\documentclass[aps,prc,preprint,amsmath,showpacs,superscriptaddress]{revtex4}
\usepackage{graphicx,epsfig}
\newcommand{\beqy}{\begin{eqnarray}}
\newcommand{\eeqy}{\end{eqnarray}}
\newcommand{\bmlet}{\begin{subequations}}
\newcommand{\emlet}{\end{subequations}}
\newcommand{\bfdel}{\mbox{\boldmath$\nabla$}}

\begin{document}

\textwidth 16.2 cm
\oddsidemargin -.54 cm
\evensidemargin -.54 cm

\def\gsimeq{\,\,\raise0.14em\hbox{$>$}\kern-0.76em\lower0.28em\hbox  
{$\sim$}\,\,}  
\def\lsimeq{\,\,\raise0.14em\hbox{$<$}\kern-0.76em\lower0.28em\hbox  
{$\sim$}\,\,}

\title{Further explorations of Skyrme-Hartree-Fock-Bogoliubov mass formulas.
IX:  Constraint of pairing force to $^1S_0$ neutron-matter gap.}
\author{N.~Chamel}
\author{S.~Goriely}
\affiliation{Institut d'Astronomie et d'Astrophysique, CP-226, Universit\'e Libre de Bruxelles, 
1050 Brussels, Belgium}
\author{J.M.~Pearson}
\affiliation{D\'ept. de Physique, Universit\'e de Montr\'eal, Montr\'eal (Qu\'ebec), H3C 3J7 Canada}
\date{\today}

\begin{abstract}
In this latest of our series of Skyrme-HFB mass models, HFB-16, we introduce 
the new feature of requiring that the contact pairing force reproduce at
each density the $^1S_0$ pairing gap of neutron matter as determined in 
microscopic calculations with realistic nucleon-nucleon forces. We retain the
earlier constraints on the Skyrme force of reproducing the energy-density curve
of neutron matter, and of having an isoscalar effective mass of $0.8M$ in
symmetric infinite nuclear matter
at the saturation density; we also keep the recently adopted
device of dropping Coulomb exchange. Furthermore, the correction term for
the spurious energy of collective motion has a form that is known to favour
fission barriers that are in good agreement with experiment. Despite the
extra constraints on the effective force, 
we have achieved a better fit to the mass data than any other mean field model, the
rms error on the 2149 measured masses of nuclei with $N$ and $Z \ge$ 8 having
been reduced to 0.632 MeV; the improvement is particularly striking for
the most neutron-rich nuclei. Moreover, it turns out that even with no
flexibility at all remaining for the pairing force, the spectral pairing gaps
that we find suggest that level densities in good agreement with experiment
should be obtained. This new force is thus particularly well-suited for 
astrophysical applications, such as stellar nucleosynthesis and neutron-star 
crusts. 

\end{abstract}


\maketitle

\section{Introduction}
\label{sect.intro}

The r-process of stellar nucleosynthesis is known to depend on the masses and 
fission barriers (among other quantities) of nuclei that are so neutron-rich 
that there is no hope of being able to measure them in the laboratory in the 
foreseeable future~\cite{agt07}. Moreover, the composition of the outer crust 
of neutron stars depends on the masses of nuclei lying as far out on the 
nuclear chart as the neutron drip line~\cite{haen07}. It is thus of the 
greatest importance to be able to make 
reliable extrapolations of these quantities away from the known region, 
relatively close to the stability line, out towards the neutron drip line. In 
order to put these extrapolations on as rigorous a footing as is feasible at 
the present time we have constructed a series of mass models based on the 
Hartree-Fock-Bogoliubov (HFB) method with Skyrme forces and a contact pairing 
force~\cite{ms02,sg02,ms03,sg03,ms04,sg05,sg06,sg07,gp08}, the parameters of
which are fitted to essentially all the available mass data. Clearly, the fit 
to the mass data must be as good as possible; hitherto our best fit was that 
of model HFB-8~\cite{ms04}, for which the rms deviation with respect to the 
2149 measured masses of nuclei with $N$ and $Z \ge$ 8 given in the most recent 
data compilation, the 2003 Atomic Mass Evaluation (AME)~\cite{awt03}, is 
0.635 MeV.  

However, masses are not the only nuclear quantity of astrophysical interest
that can be calculated with an effective interaction, and we have been paying
an increasing amount of attention to such questions as level densities, fission 
barriers and various properties of the inner crust of neutron stars
(see Section 5 of Ref.~\cite{sg05} for a brief discussion of the structure of 
neutron stars that is sufficient for our present purposes, and 
Refs.~\cite{haen07} for a more complete account).
Our ultimate aim is to construct a universal nuclear effective interaction for 
all the various astrophysical applications, and to this end we have been 
imposing on our mass models an increasing number of relevant constraints, even 
if this entails a slight deterioration in the quality of the mass fit. The 
progress made so far in this respect can be summarized as follows.

i) To make reliable calculations of the inner crust of neutron 
stars, where neutron-rich clusters coexist with a neutron liquid which may be 
superfluid, it 
is essential that the value of the effective mass implied by the effective 
interaction be realistic. For its isoscalar component $M_s^*$ the preferred
value at the equilibrium density $\rho_0$ of symmetric infinite nuclear matter
(INM) is around 0.8$M$ (see Ref.~\cite{sg07} for a summary of the experimental 
and theoretical evidence). While a phenomenological value of $M_s^*/M$ much 
closer to unity had been traditionally regarded as essential for good masses
in mean-field calculations, we showed~\cite{sg03} that it is possible to have
fits that are almost as good with $M_s^*/M$ taking the realistic value of 0.8.
The key to this success lay with exploiting the degree of freedom associated 
with the pairing ; it should nevertheless be stressed 
that reducing $M_s^*/M$ in this way still has an adverse effect on 
single-particle (s.p.) spectra in the vicinity of the Fermi level~\cite{sg03}.  

ii) Beginning with mass model HFB-9~\cite{sg05} we require that the effective
interaction reproduce the energy-density curve of neutron matter, as calculated
with realistic 2- and 3-nucleon forces~\cite{fp81}.
Not only is this condition obviously relevant to all neutron-star applications,
but it will also improve confidence in finite-nucleus extrapolations out 
towards the neutron drip line. Of all our constraints this was the one having
the most adverse effect on the mass fit, but it was still possible to hold
the deterioration to within tolerable limits.

iii) By taking a pairing force that was weaker than that of model HFB-9 and of
all earlier models we found that it was possible, for the first time with
our models, to obtain level densities in reasonable agreement with 
experiment~\cite{sg06,hg06}, while maintaining an acceptable mass fit.

iv) By making a phenomenological adjustment of the collective correction it
is possible to fit both masses and fission barriers within the same model
(HFB-14)~\cite{sg07}.

v) Our latest model (HFB-15)~\cite{gp08} incorporates the frequently 
recommended procedure of dropping Coulomb exchange, thereby simulating 
neglected effects such as Coulomb correlations, charge-symmetry breaking of 
the nuclear forces, vacuum polarization, etc. A significant 
improvement in the mass-data fit was achieved. 

As an alternative to non-relativistic Skyrme-based mass models, mass models 
based on the relativistic mean-field (RMF) method have also been proposed
\cite{lrr99,gtm05}. Now the isospin dependence of the spin-orbit field is 
{\it formally} quite different in the two approaches, and it might be thought 
that RMF models would provide more reliable extrapolations to the neutron drip 
line, since one must inevitably prefer a model that respects Lorentz covariance
to one that does not, other things being equal. However, it has been shown that
if the same mass data are fitted equally well in the two approaches then 
essentially the same predictions are made out at the neutron drip line 
\cite{pea01}. Thus we believe that the Skyrme approach is just as reliable as
the RMF method; in any case, the latter method has not yet been developed to 
the point where its mass fits are as good as those given by Skyrme-based 
methods. 

In the present paper we impose the additional constraint force of
requiring that the effective interaction reproduce as a function of density the
microscopic $^1S_0$ pairing gap of neutron matter, while retaining all the 
constraints of the earlier papers. Without this new constraint it would be 
impossible to make reliable investigations of a possible superfluid phase in 
the inner crust of neutron stars. For this purpose, we develop a scheme first 
proposed by Duguet~\cite{dg04} for constructing an effective density-dependent 
contact pairing interaction which reproduces 
\emph{any} given microscopic gap \emph{exactly}. 
Instead of parametrizing the density dependence by a simple functional form,
the strength of the contact interaction is determined at each density by 
solving directly the HFB equations in neutron matter, and requiring that 
the resulting gap be equal to the given microscopic gap at the Fermi momentum 
corresponding to that density.

In Section~\ref{sect.method} we describe the way in which we implement the HFB
method, with particular emphasis on points that we overlooked in our earlier
description in Ref.~\cite{ms02}. Section~\ref{sect.pair} explains our procedure for 
determining the strength of the pairing force, while the results of the new 
mass fit are presented in Section~\ref{sect.fits}. The two appendices
summarize some relevant features of the HFB method.

\section{The HFB method}
\renewcommand{\theequation}{2.\arabic{equation}}
\setcounter{equation}{0}
\label{sect.method}

The Skyrme force that we use in the present calculations has, as with all our 
previous HFB mass models, the conventional form
\beqy
\label{1}
v^{\rm Sky}(\pmb{r_i}, \pmb{r_j}) & = & t_0(1+x_0 P_\sigma)\delta({\pmb{r}_{ij}})
+\frac{1}{2} t_1(1+x_1 P_\sigma)\frac{1}{\hbar^2}\left[p_{ij}^2\,\delta({\pmb{r}_{ij}})
+\delta({\pmb{r}_{ij}})\, p_{ij}^2 \right]\nonumber\\
& &+t_2(1+x_2 P_\sigma)\frac{1}{\hbar^2}\pmb{p}_{ij}.\delta(\pmb{r}_{ij})\,
 \pmb{p}_{ij}
+\frac{1}{6}t_3(1+x_3 P_\sigma)\rho(\pmb{r})^\gamma\,\delta(\pmb{r}_{ij})
\nonumber\\
& &+\frac{\rm i}{\hbar^2}W_0(\mbox{\boldmath$\sigma_i+\sigma_j$})\cdot
\pmb{p}_{ij}\times\delta(\pmb{r}_{ij})\,\pmb{p}_{ij}  \quad ,
\eeqy 
where $\pmb{r}_{ij} = \pmb{r}_i - \pmb{r}_j$, $\pmb{r} = (\pmb{r}_i + 
\pmb{r}_j)/2$, $\pmb{p}_{ij} = - {\rm i}\hbar(\pmb{\nabla}_i-\pmb{\nabla}_j)/2$
is the relative momentum, and $P_\sigma$ is the two-body 
spin-exchange operator. The contact pairing force that we take here acts, as before, only between nucleons of the 
same charge state $q$ ($q = n$ or $p$ for neutron or proton, respectively) 
\beqy
\label{2}
v^{\rm pair}_q(\pmb{r_i}, \pmb{r_j})= v^{\pi\,q}[\rho_n(\pmb{r}),\rho_p(\pmb{r})]~\delta(\pmb{r}_{ij})\quad ,
\eeqy
where $v^{\pi\,q}[\rho_n,\rho_p]$ is a functional of the nucleon densities that will be specified in Section~\ref{sect.pair}. 

The only paper in which we attempt to describe the way in which we implement
the HFB method is Ref.~\cite{ms02} (Section 2). Since this account was somewhat
incomplete, and in a sense misleading, we present here a more detailed
description of what we actually did. Our original presentation~\cite{ms02} was
developed within the framework of the standard formulation of Mang~\cite{mang75} and of 
Ring and Schuck~\cite{rs80}, which is certainly well 
adapted to the use of a discrete basis such as an oscillator basis, as in the
case of our own calculations. Nevertheless, some clarification is necessary 
when this formalism is applied to effective forces of the form given in 
Eqs.~(\ref{1}) and (\ref{2}).

The standard formulation~\cite{mang75,rs80} assumes two-body forces, and
starts with the expression
\beqy\label{5}
H = \sum_{ij}t_{ij}c^\dagger_i c_j +
\frac{1}{4}\sum_{ij,kl}\bar{v}_{ij,kl}c^\dagger_i c^\dagger_j c_l c_k
\eeqy
for the Hamiltonian (see, for example, Eq. (5.25) of Ref.~\cite{rs80}).
Here we are working in a fixed basis of discrete s.p. states labelled by
$i, j$, etc., e.g., an oscillator basis, with $c^\dagger_i (c_j)$ denoting
creation (destruction) operators for real nucleons in such states (the charge
type is implicit in the label). Also we have introduced the antisymmetrized
matrix element of the two-body force
\beqy\label{6}
&\,&\bar{v}_{ij,kl} = v_{ij,kl} - v_{ij,lk}   \quad ,
\eeqy
and the matrix element $t_{ij}$ of the kinetic-energy operator 
$-\hbar^2\pmb{\nabla}^2 / 2 M_q$, denoting the nucleon mass by $M_q$. 

The Bogoliubov transformation
\beqy\label{7}
\beta^\dagger_k &=& \sum_l(U_{lk}c^\dagger_l + V_{lk}c_l) \nonumber \\
\beta_k &=& \sum_l(U^*_{lk}c_l + V^*_{lk}c^\dagger_l) 
\eeqy
then defines creation (destruction) operators $\beta^\dagger_k (\beta_k)$ of
quasiparticles as linear combinations of the creation and destruction operators
of real nucleons. The essence of the HFB method is that it takes the
vacuum state of these quasiparticles as the trial function for a variational
approximation to the energy of the nucleus. Denoting this (normalised) vacuum state by
$|\Psi>$, and noting that it will be a function of the transformation
coefficients $U_{lk}, V_{ij}$, etc., the HFB ground-state energy becomes, as 
for example in Eq. (E.20) of Ref.~\cite{rs80},
\beqy\label{8}
E_{\rm HFB} \equiv <\Psi|H|\Psi> = {\rm Tr}\left(t\rho+\frac{1}{2}\Gamma\rho-
\frac{1}{2}\Delta\kappa^*\right) \quad ,
\eeqy
in which Tr denotes the trace, $\rho$ and $\kappa$ are the so-called normal and abnormal density matrices,
respectively,
\bmlet
\beqy\label{9a}
\rho_{ij} = <\Psi|c_j^\dagger c_i|\Psi> = \sum_lV_{il}^*V_{jl} = \rho_{ji}^*
\eeqy
and
\beqy\label{9b}
\kappa_{ij} = <\Psi|c_jc_i|\Psi> = \sum_lV_{il}^*U_{jl} = - \kappa_{ji}  \quad  ,
\eeqy
\emlet 
while 
\bmlet
\beqy\label{10a}
\Gamma_{kl} = \sum_{ij}\bar{v}_{ki,lj}\,\rho_{ji}
\eeqy
and
\beqy\label{10b}
\Delta_{kl} = \frac{1}{2}\sum_{ij}\bar{v}_{kl,ij}\,\kappa_{ij}
\quad .
\eeqy
\emlet

We now notice the first departure from this standard formulation that we have
to make on account of our choice of effective forces of the form given in Eqs.
(\ref{1}) and (\ref{2}). Although Eq.~(\ref{8}) remains valid, along with
Eqs.~(\ref{6}), (\ref{7}), (\ref{9a}) and (\ref{9b}), because the force in the
particle-particle channel is different from the one used in the particle-hole 
channel we must replace Eqs.~(\ref{10a}) and (\ref{10b}), respectively, by
\bmlet
\beqy\label{11a}
\Gamma_{kl} = \sum_{ij}\bar{v}_{ki,lj}^{\rm Sky}\,\rho_{ji} + \sum_{ij}\bar{v}_{ki,lj}^{\rm Coul}\,\rho_{ji} \, ,
\eeqy
where $\bar{v}_{ki,lj}^{\rm Coul}$ are the antisymmetrized matrix elements of 
the Coulomb interaction,
and
\beqy\label{11b}
\Delta_{kl} = \frac{1}{2}\sum_{ij}\bar{v}_{kl,ij}^{\rm pair}\,\kappa_{ij} 
\quad ,
\eeqy
\emlet 
in which the choice of force in the respective channels is made explicit. This
distinction was made clear in our first HFB paper~\cite{ms02} (see Eq.~(7) of
that paper, where, however, there was an error in the subscripts of the 
expression for $\Gamma_{kp}$), but it was incorrect to imply that we could 
still take a Hamiltonian of the form~(\ref{5}) as the starting point: this 
would be possible only if no matrix element $\bar{v}_{ki,lj}$ appeared in
 both $\Gamma_{kl}$ and $\Delta_{kl}$, which is not the case. Rather, one 
should really proceed via a density-functional approach, as discussed below. 

The second departure from the standard formulation that we are
obliged to make because of our choice of effective forces was not made clear
at all in Ref.~\cite{ms02}. The HFB equations resulting from the minimization 
of the total ground-state energy~(\ref{8}) are
\beqy\label{12}
\sum_j \begin{pmatrix} h^\prime_{ij}-\lambda \delta_{ij} & \Delta_{ij} \\ -\Delta_{ij}^* & -h^{\prime\,*}_{ij} 
+ \lambda \delta_{ij} \end{pmatrix}\begin{pmatrix} U_{jk} \\ V_{jk} \end{pmatrix} =
E_k \begin{pmatrix} U_{ik} \\ V_{ik} \end{pmatrix}  \quad ,
\eeqy
where 
$E$ are the quasi-particle energies, and 
$\lambda$ is the usual chemical potential, arising as a Lagrange multiplier. 
Also
\bmlet
\beqy\label{13a}
h_{ij}^\prime = \frac{\partial\,E_{\rm HFB}}{\partial\rho_{ji}} = 
h_{ji}^{\prime\,*} \, , 
\eeqy
denoting matrix elements of the self-consistent s.p.~Hamiltonian, and
\beqy\label{13b}
\Delta_{ij} = \frac{\partial\,E_{\rm HFB}}{\partial\kappa^*_{ij}} =
-\Delta_{ji} \, ,
\eeqy
\emlet
denoting matrix elements of the pairing potential. 
Now Eq.~(\ref{12}) differs from Eq.~(6) of Ref.~\cite{ms02} in that the 
quantity $h_{ij}$ defined in the first member of Eq.~(7) of that paper, 
\beqy\label{14}
h_{ij} = t_{ij} + \Gamma_{ij} \, ,
\eeqy
has been replaced here by
\beqy\label{15}
h_{ij}^\prime = h_{ij} + h_{ij}^{\rm rear}    \, ,
\eeqy
where we have introduced matrix elements of the rearrangement s.p. field
\beqy\label{15b}
h_{ij}^{\rm rear} \equiv \frac{1}{2} \sum_{klpm}\left(
 \frac{\partial \bar{v}^{\rm Sky}_{kl,pm}} {\partial\rho_{ji}}\rho_{ml}\rho_{pk} - 
\frac{1}{2} \frac{\partial\bar{v}_{kl,pm}^{\rm pair}}{\partial\rho_{ji}}
\kappa_{pm}\kappa^*_{lk}\right)   \quad ,
\eeqy
which arises entirely from the derivative with respect to density of the
density-dependent components of the force (the Coulomb interaction is 
independent of the density and therefore does not lead to rearrangement terms).
Although these rearrangement terms 
were omitted from Eq.~(6) of Ref.~\cite{ms02}, they were included in all our
actual calculations. (The rearrangement terms were also dropped from Eq. (7.42)
of Ref.~\cite{rs80}, and from Eq. (14) of Ref.~\cite{bhr03}.
Note also that Eq.~(7.42) of Ref.~\cite{rs80} 
absorbs $\lambda$ into $h$.) Let us remark that the forces considered in this work 
depend only on the normal nucleon densities but not on the abnormal densities,
whence the expressions~(\ref{13b}) and (\ref{11b}) are equivalent. 
The foregoing formalism could certainly be implemented if one calculated the 
two-body matrix elements $\bar{v}_{ij,kl}^{\rm Sky}$ and 
$\bar{v}_{ij,kl}^{\rm pair}$ for forces (\ref{1}) and (\ref{2}), respectively, 
in the oscillator basis. However, although this presents no problem of 
principle, it is not what one does, despite suggestions to the contrary in Ref. 
\cite{ms02}. Rather, one follows  a simpler approach that does not require the 
calculation of these matrix elements, even if one still solves the HFB 
equations~(\ref{12}) in the oscillator basis, as we do here. Instead of 
starting with the 
Hamiltonian~(\ref{5}), one begins by noting that for zero-range interactions, 
such as forces~(\ref{1}) and (\ref{2}) considered in this work, the HFB energy 
can be written as the integral of a purely local energy-density functional
\beqy\label{16}
E_{\rm HFB} = \int \mathcal{E}_{\rm HFB}(\pmb{r})\,{\rm d}^3\pmb{r} \quad ,
\eeqy
where, assuming time reversibility, as always in this paper,
\beqy\label{17}
\mathcal{E}_{\rm HFB}(\pmb{r}) &=& \mathcal{E}_{\rm Sky}\Big[\rho_n(\pmb{r}),
\bfdel\rho_n(\pmb{r}), \tau_n(\pmb{r}),{\bf J}_n(\pmb{r}), \rho_p(\pmb{r}),
\bfdel\rho_p(\pmb{r}), \tau_p(\pmb{r}),{\bf J}_p(\pmb{r})\Big]  \nonumber \\
 &+&\mathcal{E}_{\rm Coul}\Big[\rho_p(\pmb{r})\Big] +
\mathcal{E}_{\rm pair}\Big[\rho_n(\pmb{r}), \tilde{\rho}_n(\pmb{r}),
\rho_p(\pmb{r}), \tilde{\rho}_p(\pmb{r})\Big] \quad ,
\eeqy
in which the usual normal density $\rho_q$, kinetic-energy density $\tau_q$,
spin-current density $\pmb{J}_q$ and pairing density $\tilde{\rho}_q$ appear
(see Appendix \ref{appA} for precise definitions).

Expressions for the first term in Eq.~(\ref{17}), the energy density 
associated with the 
Skyrme force~(\ref{1}) and implicitly including the kinetic energy, can be 
found, for example, in Ref. \cite{fpt01}. The second term in Eq.~(\ref{17})
represents the Coulomb-energy density, the exchange part of which is  
treated in our code by 
the Kohn-Sham approximation~\cite{ks65} (usually referred to as the 
Slater approximation in the nuclear-physics literature, but the original 
Slater version~\cite{sla51} is somewhat different), assuring thereby that the 
energy-density functional remains local. Actually, in the model described in
the present paper, HFB-16, we drop the Coulomb-exchange term, as discussed in
Section \ref{sect.intro}, and thus write 
\beqy\label{A3}
\mathcal{E}_{\rm Coul} = \frac{1}{2}e\rho_{\rm ch} V^{\rm Coul} \, ,
\eeqy
in which $e\rho_{\rm ch}$ is the charge density associated with protons
(this differs from $e\rho_p$ because we are taking account of the finite
size of the proton), and $V^{\rm Coul}$ is the electrostatic potential,
given by
\beqy\label{A4}
V^{\rm Coul}(\pmb{r}) = e \int d^3\pmb{r^\prime}\,
\frac{\rho_{\rm ch}(\pmb{r^\prime})}{|\pmb{r}-\pmb{r^\prime}|} \, .
\eeqy
The last term in Eq.~(\ref{17}) is the pairing-energy density, 
associated with the pairing force~(\ref{2}), and is given by 
\beqy\label{A6}
\mathcal{E}_{\rm pair}(\pmb{r})=\frac{1}{4} \sum_{q=n,p}  
v^{\pi q} [\rho_n(\pmb{r}),\rho_p(\pmb{r})] \tilde{\rho}_q(\pmb{r})^2 \,.
\eeqy

By taking a density functional of the general form given by Eqs.~(\ref{17}), 
rather than the Hamiltonian (\ref{5}), as the starting point, the possibility 
of having different equivalent forces in the particle-hole and 
particle-particle channels arises quite naturally. Indeed, in this formulation 
the forces never 
have to appear explicitly, and one even has the freedom to choose the form of
the functional in such a way that no corresponding forces exist at all. 
However, in this paper our choice of functional does correspond rigorously to 
forces of the form~(\ref{1}) and (\ref{2}), as we have already indicated. 

When the HFB energy is expressed in terms of an energy-density functional, as
in Eqs.~(\ref{16}) and (\ref{17}), then minimizing it with respect to the 
normal and pairing density matrices, for a fixed average number of neutrons and
protons, leads to the HFB equations in coordinate space, as shown by
Dobaczewski and co-workers~\cite{doba84,doba96} (the link with the 
discrete-basis formulation~(\ref{12}) is discussed at the end of 
Appendix~\ref{appA}). The advantage of solving 
the HFB equations in coordinate space, as in Refs.~\cite{doba84,doba96}, is 
that it facilitates an accurate determination of the asymptotic wavefunctions, 
but it has little impact on the determination of binding energies, which are 
our main concern. Thus for convenience we solve the HFB equations in the 
discrete-basis form~(\ref{12}). An essential step in this solution is the 
computation of the oscillator matrix elements (\ref{13a}) and (\ref{13b}), but
we considerably simplify the numerical task by generalizing to HFB the 
procedure outlined by Vautherin for the first HF-BCS calculation of deformed 
nuclei \cite{v73}. This procedure (see Appendix \ref{appA} for more details) 
works essentially in an oscillator basis but makes a detour into the 
coordinate-space representation by expressing these matrix elements as 
matrix elements of the self-consistent s.p Hamiltonian 
$h^\prime_q(\pmb{r})_{\sigma^\prime\sigma}$ and the self-consistent pairing 
field $\Delta_q(\pmb{r})$, respectively, both of which appear in the 
coordinate-space form of the HFB equations~(\ref{12b}), (explicit expressions 
for both these fields are given in Appendix \ref{appA}). In this way the 
computation of the far more numerous matrix elements (\ref{6}) of the two-body 
force is no longer required. From this point on,
our calculations proceed on the general lines described in Ref.~\cite{ms02}.

\section{Determination of density dependence of pairing force}
\label{sect.pair}
\renewcommand{\theequation}{3.\arabic{equation}}
\setcounter{equation}{0}

Little is known about the density dependence of the effective pairing 
interaction. The simplest ansatz, commonly referred as ``volume pairing'', is 
simply to assume that $v^{\pi\,q}[\rho_n,\rho_p]=V^{\pi\,q}$,
a constant, independent of the density, in which case the corresponding 
rearrangement term vanishes. This prescription has been adopted for all our 
previous forces except for the parameter sets BSk3~\cite{ms03}, 
BSk5~\cite{sg03} and BSk7~\cite{sg03}, for which we 
considered a density dependence of the usual form
\beqy\label{3}
v^{\pi\,q}[\rho_n,\rho_p] = V_{\pi q}\left\{1-\eta
\left(\frac{\rho}{\rho_0}\right)^\alpha\right\}  \quad ,
\eeqy
where $\rho=\rho_n+\rho_p$, while $V_{\pi q}$, $\eta$ and $\alpha$ are 
adjustable parameters. 
Testing different sets of parameters, it was concluded that this particular 
form~(\ref{3}) does not improve the global fit to the experimental mass 
data, and we thus abandoned it in our subsequent parametrizations. However, if 
one is to reproduce not only masses but also the microscopic pairing gaps given by bare 
nucleon-nucleon potentials, i.e., interactions fitted directly to the 2- and 
3-nucleon data, then the effective pairing force must be density dependent, as  
was first shown by Garrido \textit{et al}.~\cite{gar99} with a contact pairing 
force with the functional form~(\ref{3}). 

\begin{figure}
\centerline{\epsfig{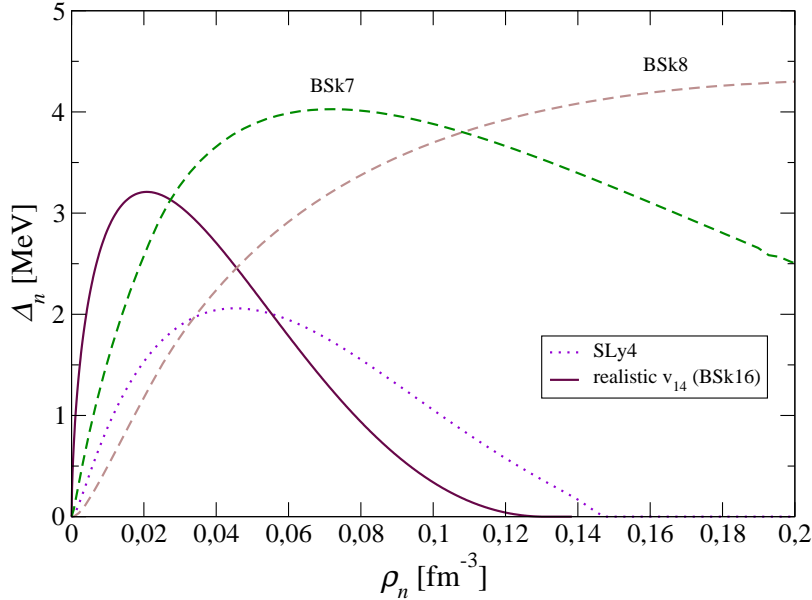}}
\caption{$^1S_0$ pairing gap $\Delta_n$ of neutron matter 
as a function of the density $\rho_n$ calculated at the Fermi momentum $k_{{\rm F}n}=(3\pi^2 \rho_n)^{1/3}$ 
for the realistic Argonne $v_{14}$ interaction~\cite{lom01}, and 
various effective forces. Note that the effective force BSk16 derived in this 
paper yields a pairing gap indistinguishable from the curve labelled ``realistic $v_{14}$ (BSk16)''.}
\label{fig_deltan_bcs}
\end{figure}

In the present paper, we wish to constrain our effective force to the pairing properties 
of pure neutron matter. Several microscopic many-body calculations of the $^1S_0$ pairing gaps 
have been carried out using various approaches and approximation schemes. Calculations 
at the simplest BCS level with the free s.p. spectrum for different nucleon-nucleon 
potentials yield very similar results~\cite{lom01,dean03, bal05}. In fact, at this level 
the pairing gap is essentially determined by the experimental $^1S_0$ phase shifts~\cite{elg98}. 
Calculations going beyond the BCS approximation generally show that medium effects lower the maximum 
of the gap function $\Delta_n(\rho_n)$ but they disagree among themselves as to the precise
density dependence. 
(Note that we use the notation $\Delta_n(\rho_n)$ for the pairing gap evaluated
at the Fermi momentum $k_{{\rm F}n}=(3\pi^2 \rho_n)^{1/3}$.) Given this lack of
agreement, we take here for the microscopic pairing gap the one shown in Fig.~7
of Ref.~\cite{lom01}, given by the realistic Argonne $v_{14}$ force in the BCS 
approximation. In any case, recent quantum Monte-Carlo 
calculations~\cite{fab05,abe07,gez07,gan08} suggest that 
the ``true'' gap might be actually close to the BCS one.

We show this microscopic BCS gap as a function of neutron-matter density in Fig.~\ref{fig_deltan_bcs}; 
for convenience, we have used the essentially exact analytical representation~\cite{kam01}
\begin{equation}
\label{eq.fit.gap}
\Delta_n(\rho_n)=\theta(k_{\rm max}-k_{{\rm F}n})\, \Delta_0 \frac{k_{{\rm F}n}^2}{k_{{\rm F}n}^2+k_1^2}\frac{(k_{{\rm F}n}-k_2)^2}{(k_{{\rm F}n}-k_2)^2+k_3^2}\, ,
\end{equation}
where $\theta$ is the Heaviside unit-step function, and the associated parameters are given in Table~\ref{tab1}. 
For comparison, we also display in the same figure the gaps corresponding to various effective interactions. 
The large differences between the neutron matter gaps
for our forces BSk7 and BSk8 is particularly striking, given that they both yield good mass fits and 
hence comparable pairing gaps in finite nuclei. The gap obtained with the force BSk7 is closer to the microscopic one, 
which might be related to the fact that its 
pairing component is density dependent, but the discrepancy with respect to the
microscopic gap is still quite unacceptable. Much better agreement
is found for the Skyrme force SLy4~\cite{cha98}, using the pairing adopted in
Ref.\cite{sto03}. However, the rms deviation of the mass fit given
by this model is very large (5.1 MeV for the subset of even-even nuclei~\cite{sto03,doba04}). 
In the present paper we show that despite the failure of the BSk7 and BSk8
models to yield acceptable pairing gaps in neutron matter, it is still possible
to construct a modified BSk model that fits the microscopic pairing gap
of neutron matter {\it exactly}, while maintaining the quality of the mass fits
obtained with our earlier models. We label the new model BSk16, as indicated
in Fig.~\ref{fig_deltan_bcs}. 

Now if we adopted the parametrization (\ref{3}) for the density dependence of 
the contact pairing force in our new mass model a serious problem would arise.
In any calculation of this sort the values of the pairing parameters depend on 
the effective interaction adopted in the particle-hole channel, essentially 
through the effective mass (see Eqns. (\ref{eq.vpi}) and (\ref{eq.rhoq}) below). In 
the case of the work of Garrido \textit{et al.}~\cite{gar99}, a
specific force (Gogny) was chosen at the outset as the interaction for the
particle-hole channel, whence the parameters $V_{\pi q}$, $\eta$ and $\alpha$ 
of the pairing force (\ref{3}) could be easily fitted 
to a microscopic pairing gap. However in the present work, the parameters of 
the Skyrme force acting in the particle-hole channel are not known 
\textit{a priori} but rather are
determined \textit{a posteriori} by fitting to experimental nuclear mass data,
as discussed in Section~\ref{sect.fits}. Fitting the parameters of the
effective interaction in both the particle-hole and particle-particle
channels, while simultaneously reproducing the microscopic pairing gap
$\Delta_q(\rho_q)$, would be an extremely onerous numerical task when using a
phenomenological functional, such as Eq.~(\ref{3}). Moreover, there is no
guarantee that the parametrization (\ref{3}) of the density dependence will be 
optimal in general, as indicated in Ref.~\cite{mar07}, 
even though it works well in the case considered by
Garrido \textit{et al.}~\cite{gar99}. Both of the above problems are avoided in the
fitting procedure that we adopted in the present paper, and now describe.

Following Duguet~\cite{dg04} and more recent calculations~\cite{mar07b}, we 
assume that the pairing force for nucleons $q$ depends only on $\rho_q$,
and not on the total density $\rho$. Besides, we suppose that the effective 
interaction $v^{\pi\,q}[\rho_q(\pmb{r})]$ at the point $\pmb{r}$ is the same 
as that in INM of pure nucleon species $q$ at the density $\rho_q(\pmb{r})$. 
But instead of postulating a simple functional form for the
density dependence, we have determined the strength of the effective pairing 
force $v^{\pi\,q}[\rho_q]$ at each nucleon density $\rho_q$ by solving the 
HFB equations in uniform matter and requiring that the resulting gap
reproduce \emph{exactly} the given microscopic pairing gap $\Delta_q(\rho_q)$ at that density. 
The only free parameter is the energy cutoff $\varepsilon_{\Lambda}$. Note that in the present work, 
we have set an upper limit on the s.p. states that are retained in the spectrum 
without using the Bulgac-Yu~\cite{by02} regularization procedure, which we adopted 
in models HFB-12 to HFB-15~\cite{sg06,sg07,gp08}. 

In uniform matter, the HFB equations reduce to the BCS equations~(\ref{C7}) and 
(\ref{C8}). Solving these equations for $v^{\pi\,q}[\rho_q]$ in pure matter of 
nucleon species $q$ and adopting the cutoff prescription (iii) discussed in Appendix~\ref{appC}, 
yields (after a change of variable $\xi = \varepsilon - U_q$ in the integrals)
\beqy
\label{eq.vpi}
v^{\pi\,q}[\rho_q]=-8\pi^2\left(\frac{\hbar^2}{2 M_q^*(\rho_q)}\right)^{3/2} 
\left(\int_0^{\mu_q+\varepsilon_{\Lambda}}{\rm d}\xi 
\frac{\sqrt{\xi}}{\sqrt{(\xi-\mu_q)^2+\Delta_q(\rho_q)^2}}
\right)^{-1}  \quad ,
\eeqy
where $M_q^*(\rho_q)$ is the effective nucleon mass in matter of pure nucleon 
species $q$ at density $\rho_q$, as given by Eq.~(\ref{A7}), 
and $\mu_q\equiv\lambda_q - U_q$ can be obtained by solving
\beqy
\label{eq.rhoq}
\rho_q = \frac{1}{4\pi^2} \left(\frac{2 M_q^*(\rho_q)}{\hbar^2}\right)^{3/2} \int_0^{\infty}{\rm d}\xi \sqrt{\xi}\left(1-\frac{\xi-\mu_q}{\sqrt{(\xi-\mu_q)^2+\Delta_q(\rho_q)^2}}\right) \, .
\eeqy
Note that determining the strength $v^{\pi\,q}$ of 
the effective pairing force by Eq.~(\ref{eq.vpi}) ensures its automatic renormalization for any 
changes of the energy cutoff $\varepsilon_{\Lambda}$. The same value of $\varepsilon_\Lambda$ is used 
in calculations of finite nuclei (see Eq.~(\ref{4.3}) below). 
Let us remark that if, instead of imposing a cutoff 
above the Fermi level, we would have chosen a fixed cutoff (as was done in 
model HFB-1~\cite{ms02}), Eq.~(\ref{eq.vpi}) would have been 
actually an integro-differential equation for $v^{\pi\,q}[\rho_q]$ as discussed in Appendix~\ref{appC}. 
In such case, it would therefore have been much more difficult to compute the strength of the 
effective pairing force. 

In contrast to the use of parametrized pairing forces~(\ref{3}), in the present scheme 
we reproduce \emph{exactly} the density dependence of the microscopic pairing gap. 
At the same time, the fitting procedure of the Skyrme parameters is greatly simplified: 
at each iteration of the mass fit, the pairing function $v^{\pi\,q}[\rho_q]$
 is determined unambiguously through the coupled equations~(\ref{eq.vpi}) and (\ref{eq.rhoq}) 
by the given microscopic gap $\Delta_n(\rho_n)$ 
and the running values of the Skyrme parameters $t_1$, $t_2$, $x_1$ and $x_2$.

\begin{figure}
\centerline{\epsfig{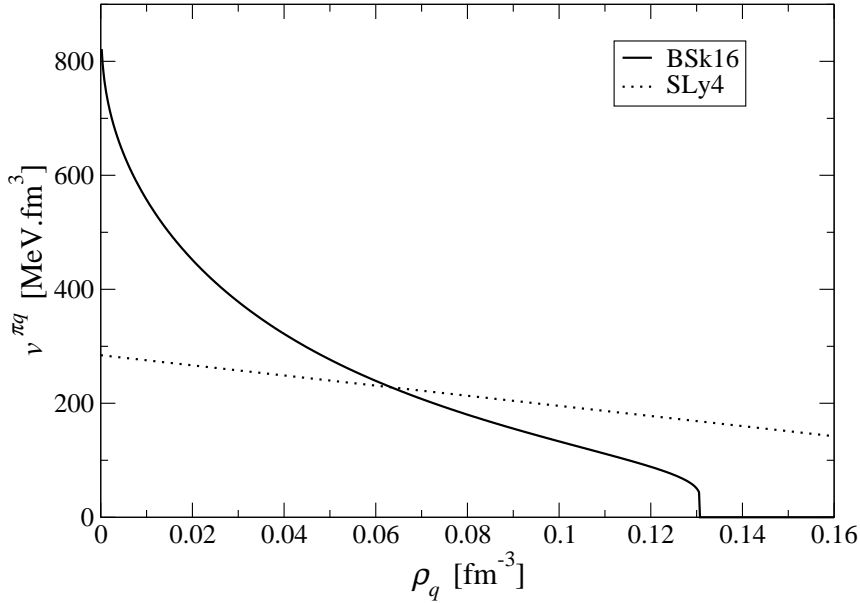}}
\caption{Density dependence of the strength of the effective pairing force 
$v^{\pi\,q}[\rho_q]$, which has been fitted to 
the $^1S_0$ pairing gap $\Delta_n$ shown in Fig.~\ref{fig_deltan_bcs}. 
The effective mass at each density is calculated with the final BSk16 Skyrme 
parameters (see Section~\ref{sect.fits}). For comparison, we show the effective 
pairing force used in the SLy4 mass model~\cite{sto03,doba04}.}
\label{fig_vpiq}
\end{figure}

In our actual numerical calculations, instead of solving both 
Eqs.~(\ref{eq.vpi}) and (\ref{eq.rhoq}) at each density, we approximate the
reduced chemical potential $\mu_q$ by the Fermi energy
\beqy
\varepsilon^{(q)}_{{\rm F}} = \frac{\hbar^2 k_{{\rm F}q}^2}{2 M_q^*} \, ,
\eeqy
where $k_{{\rm F}q}$ is the Fermi wave number
\beqy
k_{{\rm F}q}= (3\pi^2 \rho_q)^{1/3} \, .
\eeqy
We have found that this approximation holds well, provided 
$\Delta_q \ll \varepsilon^{(q)}_{{\rm F}}$. 
Then we do not have to solve Eq. (\ref{eq.rhoq}) at all, and the
evaluation of $v^{\pi\,q}[\rho_q]$ through Eq.~(\ref{eq.vpi}) reduces to
a single, simple integration. The effective pairing strength 
$v^{\pi\,q}[\rho_q]$ determined for the final Skyrme force BSk16 (see Section~\ref{sect.fits})
through Eq.~(\ref{eq.vpi}), is shown in Fig.~\ref{fig_vpiq}; it will be seen that the 
density dependence is qualitatively similar to that given by Eq.~(\ref{3}). 

The foregoing neutron-matter prescription suffices in principle to fix
the neutron pairing (see, however, Section \ref{sect.fits} for the distinction that 
we make between even-$N$ and odd-$N$ nuclei). Because of charge-symmetry
breaking, the proton pairing force should in principle be determined in an
analogous way on the basis of ``proton-matter" calculations, which are
performed in exactly the same way as are neutron-matter calculations, using the
appropriate
proton-proton interaction, with, of course, the Coulomb force suppressed, since
``proton-matter'' calculations would otherwise diverge. However, this 
prescription will not suffice, since in finite nuclei the Coulomb force 
{\it does} act, and may be expected to modify the proton pairing, in a highly
complicated way. For simplicity, we take $v^{\pi\,p}[\rho_p]$ to be given by
$v^{\pi\,n}[\rho_p]$, multiplied by a constant, density-independent factor
which is the same for all nuclei and is taken as a fitting parameter (see
Section~\ref{sect.fits} for more details). 

\section{The mass fits}
\label{sect.fits}
\renewcommand{\theequation}{4.\arabic{equation}}
\setcounter{equation}{0}

\begin{figure}
\centerline{\epsfig{figure=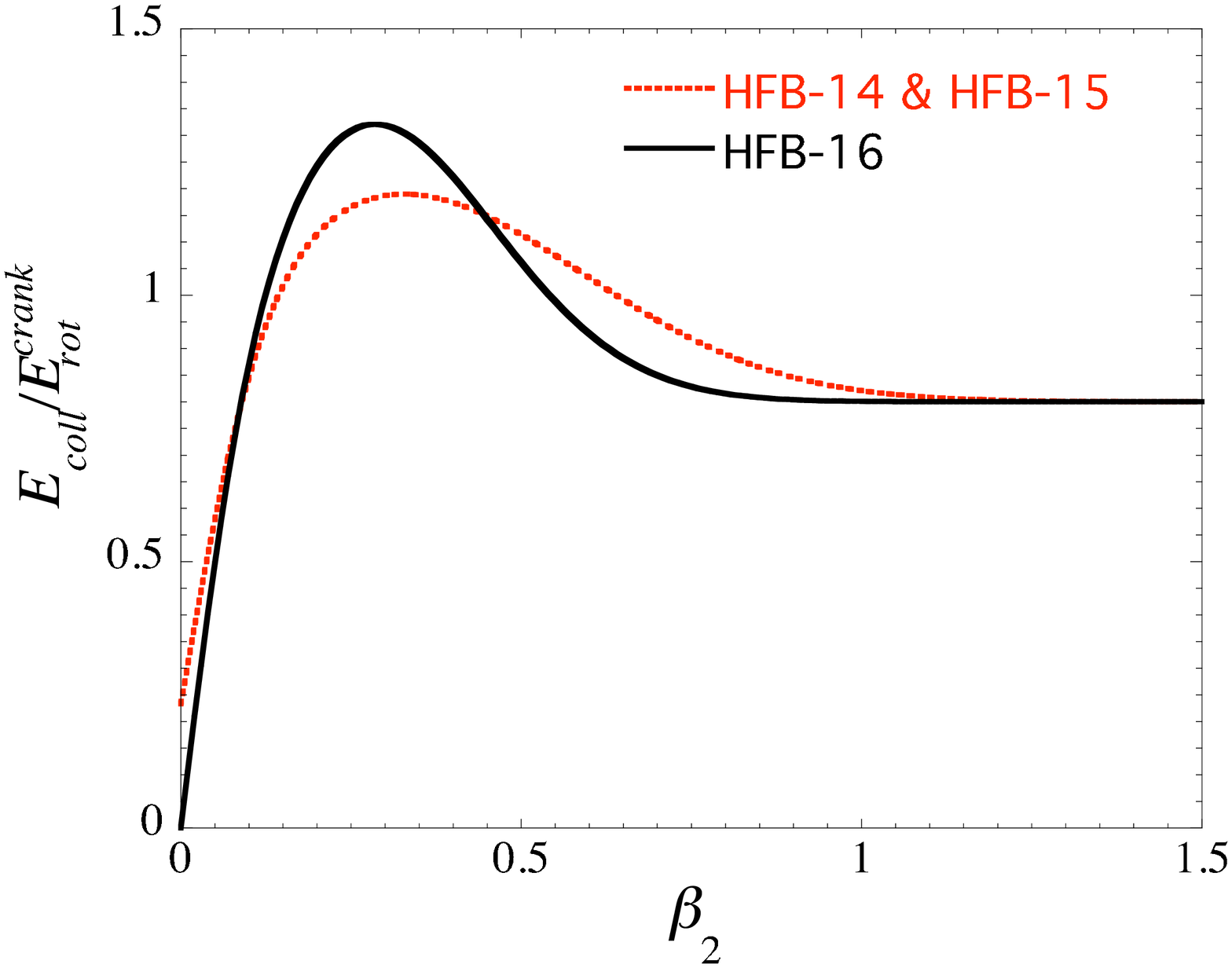,height=10.0cm}}
\caption{Deformational variation of collective correction of Eq. (\ref{4.2}).}
\label{fig_erot}
\end{figure}

We have described our treatment of the Skyrme and pairing forces of our new 
model, HFB-16, in Sections~\ref{sect.method} and \ref{sect.pair}, respectively, but, 
before presenting the parameter sets for these forces that emerge from the fits
to the mass data, we discuss several further points, as follows. 

i) {\it Wigner correction.} To the HFB energy calculated for the Skyrme force
(\ref{1}) and the pairing force~(\ref{2}) has to be added the Wigner correction
\beqy\label{4.1} 
E_W = V_W\exp\Bigg\{-\lambda\Bigg(\frac{N-Z}{A}\Bigg)^2\Bigg\}
+V_W^{\prime}|N-Z|\exp\Bigg\{-\Bigg(\frac{A}{A_0}\Bigg)^2\Bigg\} \quad ,
\eeqy
which contributes significantly only for light nuclei~\cite{sg02}. Our 
treatment of this
correction is purely phenomenological, with the first term believed to be 
representing a $T$ = 0 $n-p$ pairing~\cite{sw97,sat97,sw00}, while the second 
is characteristic of Wigner's supermultiplet theory~\cite{wig37}, based on 
SU(4) spin-isospin symmetry, although it can also be interpreted as 
corresponding to $T$ = 1 $n-p$ pairing~\cite{sw97}.

ii) {\it Collective correction.} As in all our previous models we subtract
from the calculated HFB energy an estimate for the spurious collective energy.
The form we adopt here for this correction is
\beqy\label{4.2}
E_{coll}= E_{rot}^{crank}\Big\{b~\tanh(c|\beta_2|) +
d|\beta_2|~\exp\{-l(|\beta_2| - \beta_2^0)^2\}\Big\} \quad  ,
\eeqy
in which $E_{rot}^{crank}$ denotes the cranking-model value of the rotational
correction and $\beta_2$ the quadrupole deformation, while all other parameters
are free fitting parameters. This correction is intended to take account of
both rotational and vibrational spurious energy, but while this expression for
$E_{coll}$ is seen to vanish for spherical nuclei, we know that a vibrational 
correction is still required for such nuclei. Thus we must suppose that the
vibrational correction for spherical nuclei is absorbed into the fitted force
parameters, so that it is only the deformational variation of the vibrational
correction that is represented by Eq.~(\ref{4.2}), along with the complete
rotational correction. Our final mass fit gives for the coefficient $b$ 
appearing in the first term of Eq.~(\ref{4.2}) the value of 0.8 (see Table 
\ref{tab3}), the same value as we found for model HFB-14~\cite{sg07}. Thus
the argument that was made in Ref.~\cite{sg07} points once again to the
first term of Eq.~(\ref{4.2}) being identified with the rotational
correction. 

Actually, the second term of the correction~(\ref{4.2}) differs from that of
model HFB-14~\cite{sg07} in that it carries an additional factor $|\beta_2|$, 
the role of which is to ensure that this term, like the first term, vanishes
as sphericity is approached. Some
instabilities that were encountered for spherical or weakly deformed nuclei
with HFB-14~\cite{sg07} are thereby avoided, and this modification is in 
part reponsible for the improved mass fit of this new model (see below in this
Section). The corrections for the two models are compared graphically in
Fig.~\ref{fig_erot}, where it will be seen that except in the spherical
limit the differences between the two corrections are small, especially for
large deformations. This point is of crucial importance for fission barriers,
which are highly sensitive to the collective correction at large deformations, 
and we do not expect 
that the barriers to be calculated (in a future paper) with the present
model will be appreciably different from those found with model HFB-14. In
any case, it will be possible to perform any fine tuning of the barriers that 
may be required by adjustment of the collective correction alone, without any
modification of the Skyrme or pairing parameters, our experience in 
Ref.~\cite{sg07} showing that the perturbation of the mass fit will be
minimal, provided the collective contribution has the form (\ref{4.2}). 

iii) As with models HFB-14~\cite{sg07} and HFB-15~\cite{gp08}, we impose on the
parameters of the Skyrme force the condition $M_s^*/M$ = 0.8 for the isoscalar
effective mass in symmetric INM at the density $\rho = \rho_0$ (see
Section~\ref{sect.intro}).

iv) By varying the exponent $\gamma$ the incompressibility coefficient $K_v$ 
can be varied, and we find that excellent mass fits can be obtained over
the range 230 $ \le K_v \le $ 270 MeV. However, measurements of the breathing
mode restrict this range to 230 $ \le K_v \le $ 250 MeV~\cite{col04,piek05},
so we limit the value of $\gamma$ accordingly.

v) The measured rms charge radius $R_c$ of $^{208}$Pb, 5.501 $\pm 0.001$ fm
\cite{ang04}, was required to be well reproduced; relaxing this condition leads
to very little improvement in the mass fit.

vi) As with all our models since HFB-9~\cite{sg05}, we constrain our Skyrme 
force to fit neutron matter (see Section~\ref{sect.intro}); this turns out to be
equivalent to setting the INM parameter $J$ to be 30 MeV.

vii) We drop the Coulomb-exchange term from the HFB energy-density functional
(see Section~\ref{sect.intro}). 

viii) As with all our models, we treat the case of odd $N$ and/or odd $Z$ 
by the ``level-filling" approximation. That is, the odd nucleon is placed 
with equal probability in each of the lowest-energy available degenerate
states generated by the HF calculation with the next lowest even number of 
nucleons, and then applying blocking.

ix) Our new prescription for the pairing strength, described in Section
\ref{sect.pair}, is strictly valid only for neutrons. Because of Coulomb 
effects, and a possible charge-symmetry breaking of nuclear forces, 
we must allow for the proton pairing strength to be different. Likewise, we 
shall allow the pairing to be different for odd-$A$ and odd-odd nuclei to 
compensate for our failure for such nuclei to project out states that respect 
time-reversal invariance. We take account of these extra degrees of freedom by 
multiplying the value of $v^{\pi\,q}[\rho_q]$, as determined through
Eq.~(\ref{eq.vpi}), by renormalizing factors $f^{\pm}_q$, where $f^+_p, f^-_p$ 
and $f^-_n$ are free, density-independent parameters to be included in the mass
fit; in keeping with the spirit of this paper we set $f^+_n$ = 1.  

x) In calculations of finite nuclei, we use the smooth pairing cutoff factor
\beqy
\label{4.3}
f_i = \biggl[1+\exp\left( (\varepsilon_i-\lambda-\varepsilon_\Lambda)/\tau\right) \biggr]^{-1/2}
\eeqy
with $\tau=0.25$ MeV, in order to prevent an unphysical selection of nearly 
degenerate levels whenever the cutoff energy lies between the levels. 

xi) The spurious centre-of-mass energy is removed following the essentially 
exact procedure described in Ref.~\cite{sg03}. 

xii) A correction for the finite size of the proton is made to both the charge
radius and the energy, as in all our previous HFB models. We assume a Gauss
distribution of charge over the proton, with an rms radius of 0.895 
fm~\cite{sic03}; the folding of this charge distribution over the HFB
distribution of point protons is performed using Eqs. (4.2) and (4.3) of
Ref.~\cite{neg70}. 

xiii) As in all our papers except Ref. \cite{ms04} we make no attempt to
project out states of good particle number. 

xiv) The oscillator basis in terms of which the HFB wavefunctions are expanded
contains 20 major shells.

The new force parameters, labelled BSk16, are fitted to the same set of 2149 
measured masses~\cite{awt03} as are all our models since HFB-9~\cite{sg05}; the
resulting values are shown in Table~\ref{tab2}, while Table~\ref{tab3} shows
the parameters of the collective correction of Eq.~(\ref{4.2}). These two tables
define the HFB-16 model, with which we have constructed a complete mass table 
running from one drip line to the other over the range $Z$ and
$N \ge$ 8 and $Z \le$ 110. 

The first ten lines of Table~\ref{tab2} show the Skyrme parameters; their only
noteworthy feature is the small value of 
$t_2$ and the large value of $x_2$, which implies that the $^1P$ and $^3P$ 
interactions have strengths of comparable magnitude but opposite sign, the 
latter being attractive. The next three lines show the factors by which the 
pairing strength determined from neutron matter must be renormalized for
protons and for odd nuclei. With $f_n^+ = 1$ we see that to within the limits
of numerical accuracy $f_n^-/f_n^+ = f_p^-/f_p^+$ and the proton and neutron
pairing strengths are effectively equal. This means that these three degrees of
freedom could have been reduced to a single one, the ratio of the odd-number 
pairing strength to the even-number pairing strength (as in all our previous 
models, pairing is always a little stronger for an odd number of nucleons).
Line 14 shows the pairing cutoff parameter $\varepsilon_{\Lambda}$; the value
of 16 MeV was chosen on the basis of past experience~\cite{sg06}.
The last four lines give the Wigner parameters of Eq.~(\ref{4.1}).

The rms and mean (data - theory) values of the deviations between the measured
masses and the HFB-16 predictions are given in the first and second lines,
respectively, of Table~\ref{tab4}, where we also compare with HFB-15 
\cite{gp08}, HFB-14~\cite{sg07} and our previous ``best-fit'' model HFB-8 
\cite{ms04}. We see that with this new model we have achieved our best fit
ever ($\sigma=0.632$ MeV), in addition to having a greater conformity to physical reality than with 
any of our other models. It should be recalled that the first physical 
constraint we imposed on our mass fits, conformity to the neutron-matter 
energy-density curve, led, with model HFB-9~\cite{sg05}, to a serious loss in 
the quality of the mass fit. However, we have now entirely recovered the precision of model 
HFB-8; this could be a result of the device of dropping Coulomb exchange (introduced in model 
HFB-15~\cite{gp08}), of modifying the collective correction discussed above, or of 
requiring a better agreement with the BCS neutron-matter pairing gap. Concerning this last point, 
we should emphasize that starting from a neutron-matter gap that includes medium effects 
beyond the BCS approximation, we were not able to obtain a good mass fit. 
Fitting masses to such microscopic gaps seems to be very difficult in the present framework. 

The superiority of the new model becomes still more striking on looking at
the next six lines of Table~\ref{tab4}. Lines 3 and 4 show the rms and mean 
deviations for the astrophysically crucial subset of the mass data consisting 
of the 185 neutron-rich nuclei having a neutron-separation energy 
$S_n \le$ 5.0 MeV. The following four lines give the rms and mean deviations
for the $S_n$ and the beta-decay energies $Q_{\beta}$, using the full data set 
of 2149 measured masses: since these are differential quantities they are
astrophysically more relevant than the absolute masses $M$. It is noteworthy
that in all these important categories of mass-related quantities model HFB-16 
out-performs all our other HFB models. 

Turning to charge radii, lines 9 and 10 show the rms and mean deviations 
between the measured values~\cite{ang04} and the model predictions. Only here
do we see any deterioration with respect to model HFB-8; the mean deviation
$\bar{\epsilon}(R_c)$ suggests that from model HFB-14 onwards we should have
been taking a slightly larger density $\rho_0$, but we do not know whether
this would have an adverse effect on the mass fit.

\begin{figure}
\begin{center}
\centerline{\epsfig{figure=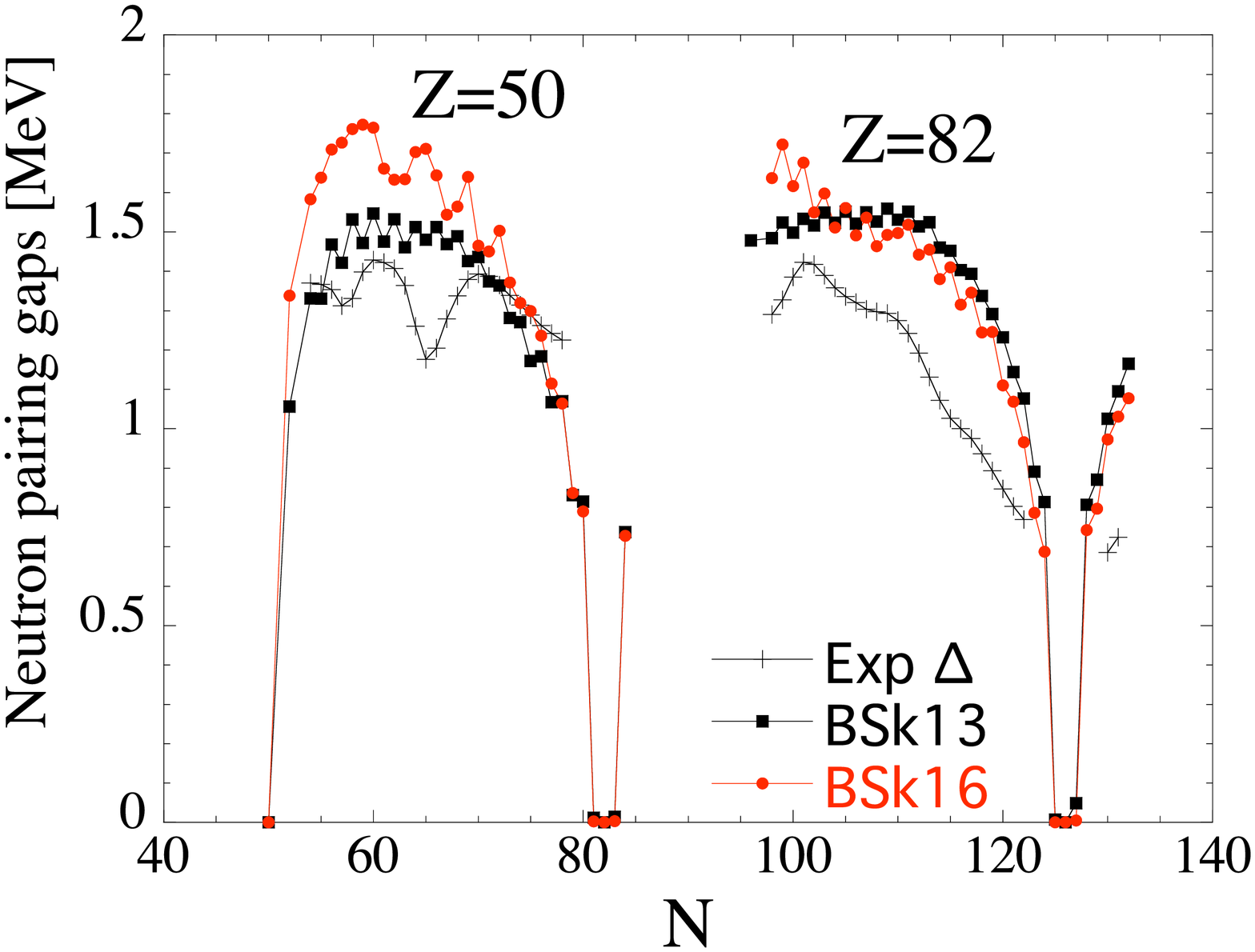,height=9.0cm,width=13cm}}
\caption{Comparison of the experimental even-odd differences $\Delta^{(5)}$
with the HFB-16 theoretical neutron spectral pairing gaps
$\langle u v\Delta\rangle$ for the Sn and Pb isotopic chains.}
\label{fig_deltan}
\end{center}
\end{figure} 

As stated in Section~\ref{sect.intro}, one of our concerns has been to have a 
pairing force that not only permits a good mass fit but also is consistent
with level densities that are in reasonable agreement with experiment. 
However, in the present model the pairing force is fixed almost entirely 
by the {\it a priori} calculated pairing properties of neutron matter, the 
flexibility offered by the renormalization factors $f^{\pm}_q$ being negligible
in this respect. Now previous experience with model HFB-13~\cite{sg06,hg06} 
shows that reasonable level densities are found when the spectral pairing gap 
$\langle u v\Delta\rangle$ of the model (defined as in Eq. (5) of 
Ref.~\cite{sg06}) lies close to the fifth-order experimental even-odd mass 
differences $\Delta^{(5)}$. We compare in Fig.~\ref{fig_deltan} the HFB-16 
spectral pairing gaps for the Sn and Pb isotope chains with those of model 
HFB-13~\cite{sg06}. It will be seen that the new model HFB-16 agrees closely 
with the earlier one in the case of the Pb isotopes, and is not substantially
stronger in the case of the Sn isotopes (similar conclusions hold for the 
third-order even-odd differences $\Delta^{(3)}$). We leave for a later paper the actual 
level-density calculations with HFB-16, but Fig.~\ref{fig_deltan} is 
encouraging. 

\begin{figure}
\begin{center}
\centerline{\epsfig{figure=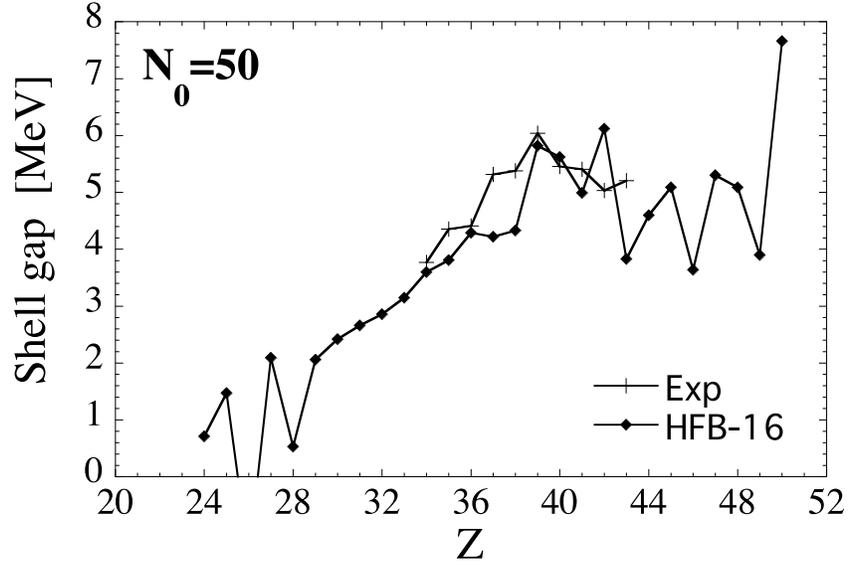,width=12cm}}
\caption{$N_0=50$ shell gap as function of $Z$ for mass model HFB-16.}
\label{fig_gap50}
\end{center}
\end{figure}

\begin{figure}
\begin{center}
\centerline{\epsfig{figure=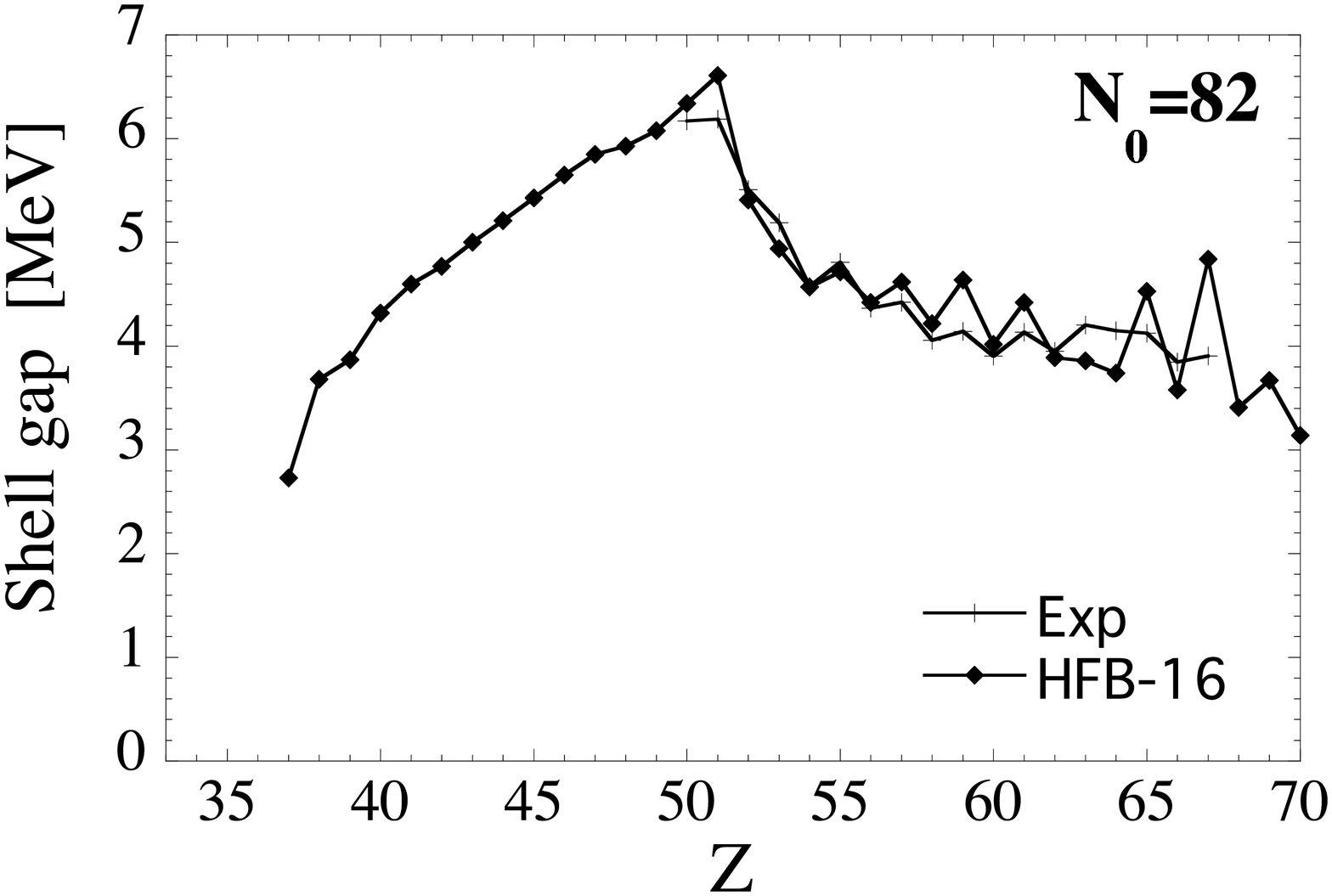,width=12cm}}
\caption{$N_0=82$ shell gap as function of $Z$ for mass model HFB-16.}
\label{fig_gap82}
\end{center}
\end{figure}

\begin{figure}
\begin{center}
\centerline{\epsfig{figure=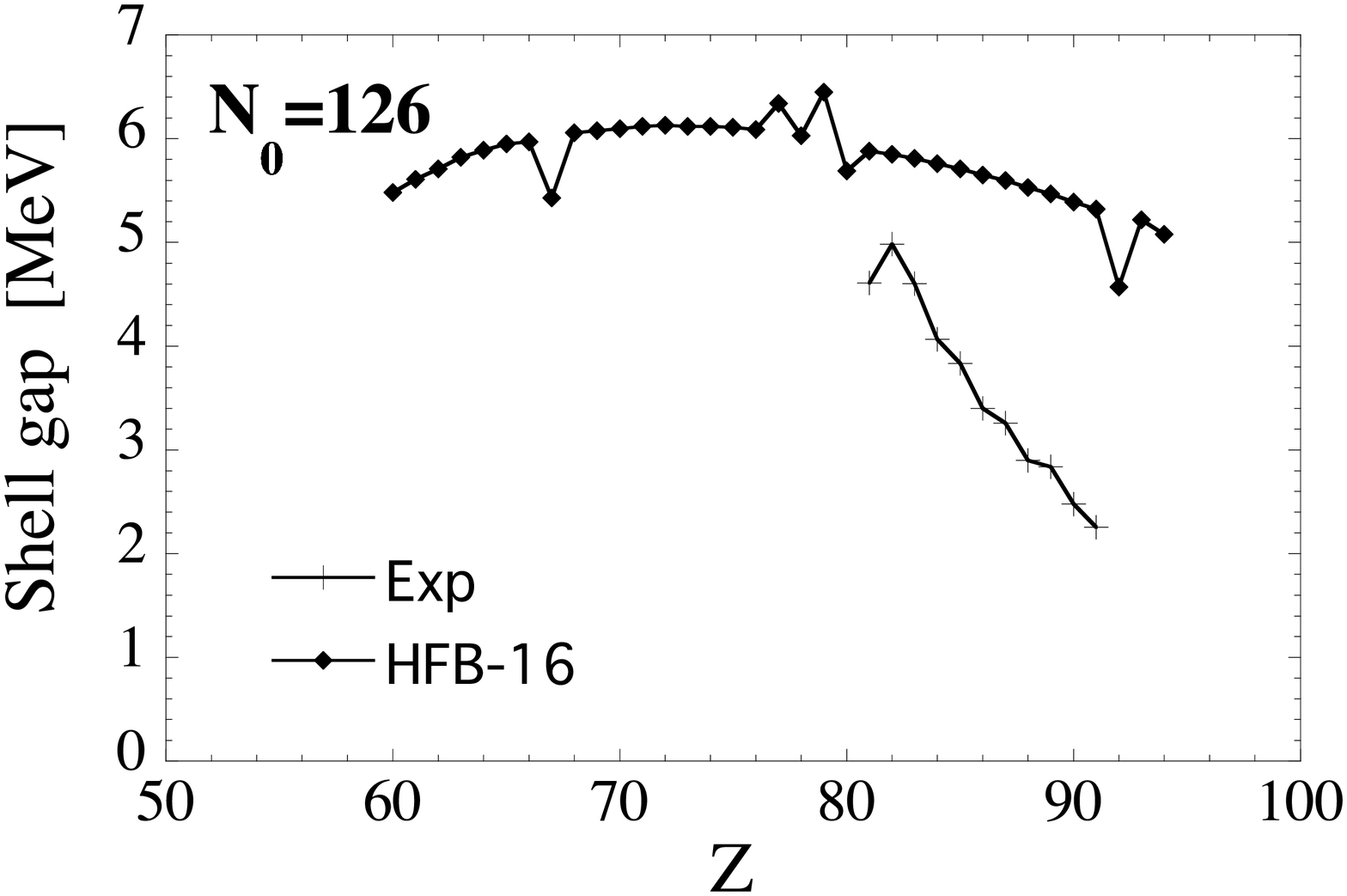,width=12cm}}
\caption{$N_0=126$ shell gap as function of $Z$ for mass model HFB-16.}
\label{fig_gap126}
\end{center}
\end{figure}

\begin{figure}
\begin{center}
\centerline{\epsfig{figure=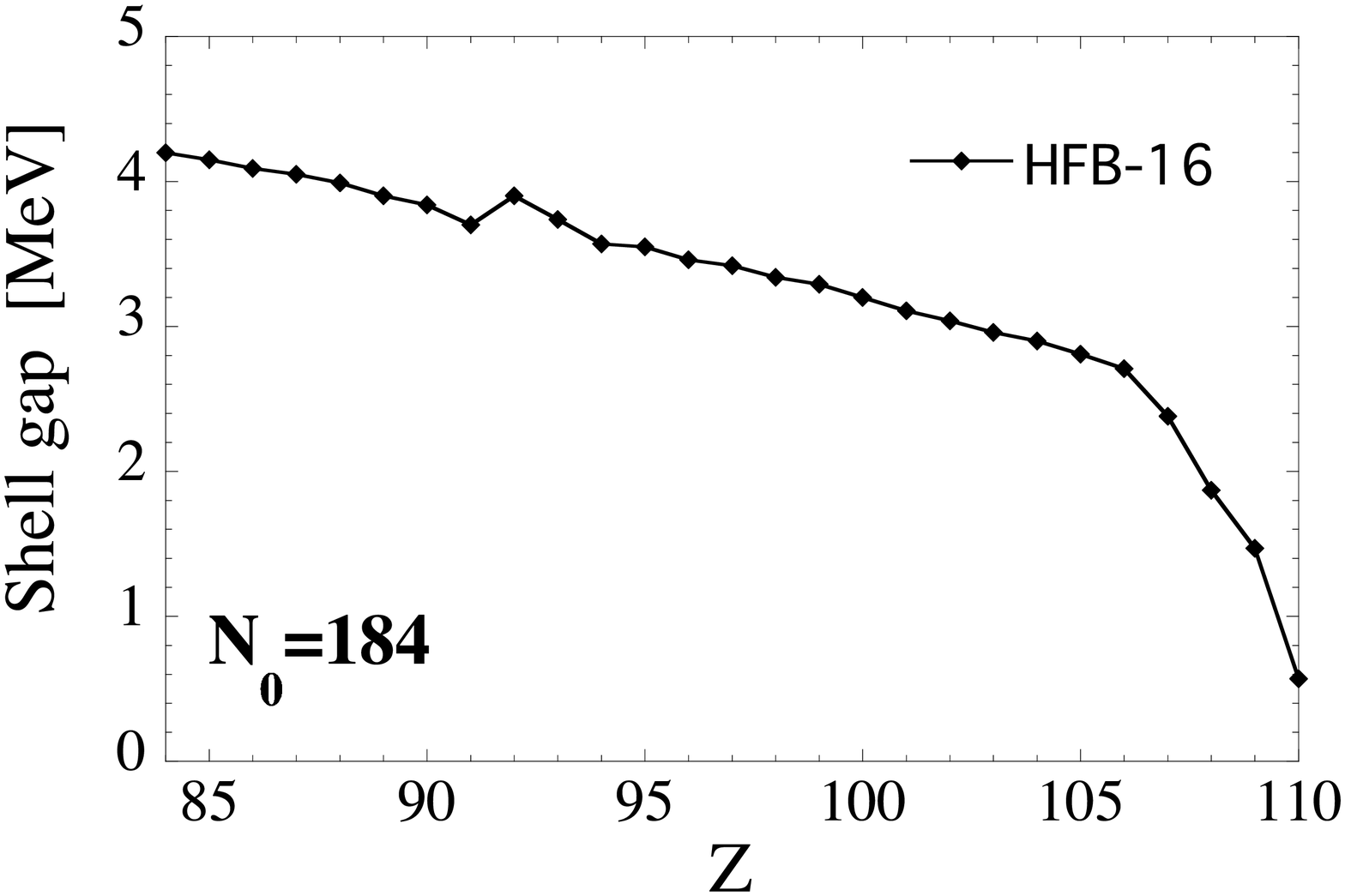,width=12cm}}
\caption{$N_0=184$ shell gap as function of $Z$ for mass model HFB-16.}
\label{fig_gap184}
\end{center}
\end{figure}

{\it Magic neutron-shell gaps.} While the global fit to the mass data given by
model HFB-16 is seen to be excellent, a few individual predictions could still
be quite anomalous without having a serious impact on the global fit. Of
particular importance in this respect are the masses involved in the definition
of the neutron-shell gaps,
\beqy
\label{shell}
\Delta_n(N_0, Z) = S_{2n}(N_0,Z) - S_{2n}(N_0+2,Z)  \quad ,
\eeqy
In Figs.~\ref{fig_gap50}-- \ref{fig_gap184} we show these gaps
as a function of $Z$ for the magic numbers $N_0 = 50, 82, 126$ and $184$. The
agreement with experiment is excellent for $N_0 = 50$ and 82, with strong
gap quenching predicted as the neutron drip line is approached. On the other
hand, for $N_0 = 126$ the agreement with experiment is quite bad, as is always
the case with our HFB models. No quenching is predicted for either $N_0$ =
126 or 184.

{\it Macroscopic properties.} Table~\ref{tab5} shows the parameters of infinite
and semi-infinite nuclear matter (INM and SINM) that we have calculated for the
force BSk16 (for the definition of these parameters see, for example,
Ref.~\cite{sg06}). Our SINM
calculations, which do not take pairing into account, are performed using the
HF code of M. Farine, as described in Appendix A of Ref.~\cite{ms05}. It should
be noted that the values of $M_s^*$ and $J$ were imposed, as described at the
beginning of this Section. In this table we show also the corresponding
parameters for our last two forces; it will be seen that there has been very
little change, although it is interesting to note that with the omission of
Coulomb exchange (BSk15 and BSk16) the binding energy per nucleon of
symmetric INM, $a_v$, has increased. All our values of these parameters are
compatible with all the available experimental data, as discussed in Ref.~\cite{sg06}.
To this earlier discussion we would like to append the remark that with all
our forces we find an isovector effective mass $M^*_v $ that is smaller than
the isoscalar effective mass $M^*_s$ at the density $\rho_0$. This result implies that the 
neutron effective mass $M^*_n$ is larger than the proton effective mass $M^*_p$ in 
neutron-rich matter. Such an isovector splitting of the effective mass is consistent 
with measurements of isovector giant resonances~\cite{les06}, and has been confirmed in several many-body
calculations with realistic forces~\cite{van05,zuo06}.

The energy-density curve of neutron matter for force BSk16 is indistinguishable 
from the realistic curve of Ref.~\cite{fp81} up to the supernuclear density of
0.3 neutron.fm$^{-3}$, as is the case with all our other forces that have been
fitted to $J$ = 30 MeV, i.e, BSk9 and all later forces (see Fig. 13 of
Ref.~\cite{sg06}). It is to be noted that unlike Ref.~\cite{les06} we have not
had to resort to a second $t_3$ term in the Skyrme force in order to
simultaneously fit neutron matter and obtain the correct sign for the
isovector splitting of the effective mass.

\begin{figure}
\begin{center}
\centerline{\epsfig{figure=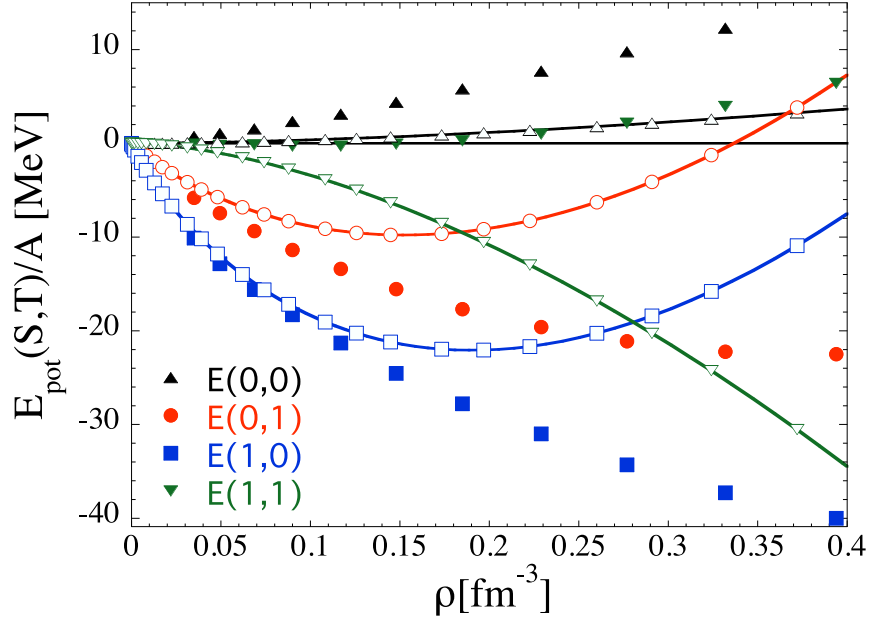,width=12cm}}
\caption{Potential energy per particle in each $(S,T)$ channel as a function of
density for symmetric INM. The full symbols correspond to BHF calculations and 
the open symbols (connected with solid lines) to the BSk16 force.}
\label{fig_inm}
\end{center}
\end{figure}

Fig.~\ref{fig_inm} shows the potential energy per particle in each of the four
two-body spin-isospin $(S,T)$ channels as a function of density for symmetric
INM; we give results for both BSk16 and Brueckner-Hartree-Fock (BHF)
calculations with realistic two- and three-nucleon forces. These latter results
are taken from Fig. 6 of Ref.~\cite{les06} and quoted as Ref. 62 of that paper.
There is reasonable agreement between BSk16 and the realistic calculations in 
all states except the (1,1) state, where BSk16 is strongly attractive,
while BHF is very weak. It seems to be very difficult, in fact, to fit both odd
states with a conventional Skyrme force. For example, SLy4~\cite{cha98} handles
the (1,1) state rather well, having imposed $x_2$ = -1, but as a result works
much worse than BSk16 for the (0,0) state. In any case,
the excessive attraction that BSk16 gives in the (1,1) state leads
to the onset of a ferromagnetic instability in neutron matter at relatively low
density, as seen from the value of $\rho_{frmg}/\rho_0$ in Table~\ref{tab5}. 

{\it Stability of extrapolation.} Although all our models agree fairly closely 
in the known region of the nuclear chart, simply by virtue of their fit to the 
data, there is no {\it a priori} guarantee that
they will give comparable extrapolations to the astrophysically interesting
region close to the neutron drip line. Accordingly, in Table~\ref{tab6} we
compare the predictions made by HFB-16 with our earlier models HFB-15, HFB-14 
and HFB-8 for all those nuclei with $26 \le Z \le 110$ for which 
$S_n <$ 4.0 MeV. This table shows that for the more astrophysically relevant 
quantities $S_n$ and $Q_\beta$ the differences between the various model 
predictions are much smaller than for the absolute masses, and in fact are 
comparable to the deviations between each model and the data. This stability
of the predictions against slight changes in the model is consistent with an
overall reliability of the HFB approach, although insofar as differences
between the model predictions are significant, we would prefer HFB-16, given 
both the excellence of its fit to the data and the high degree of reality built
into the model. In this latter respect, the conformity of both the Skyrme
and pairing components of the effective force to realistic neutron-matter
calculations should particularly enhance our confidence in predictions for
highly neutron-rich nuclei.

\section{Conclusions}

The new feature that we introduce here in this latest of our series of
Skyrme-HFB mass models, HFB-16, is the requirement that the contact pairing 
force reproduce exactly at
each density encountered in the nuclear system (nucleus or the inner crust of 
neutron stars) the $^1S_0$ pairing gap of neutron matter, as determined in 
microscopic calculations with realistic nucleon-nucleon forces. We retain the 
earlier constraints on the Skyrme force of reproducing the energy-density curve
of neutron matter, and of having an isoscalar effective mass of $0.8M$ in 
symmetric
INM at the saturation density $\rho_0$; we also keep the recently adopted
device of dropping Coulomb exchange. Furthermore, the correction term for
the spurious energy of collective motion has a form that is known to favour
fission barriers that are in good agreement with experiment.

Remarkably, despite the severe constraints imposed by neutron matter on both 
the Skyrme and pairing components of the effective force we have achieved the 
best fit ever to the mass data within the framework of mean-field models (the 
improvement is particularly striking for the most neutron-rich nuclei).
The rms error $\sigma$ of the 2149 measured masses of nuclei with $N$ and 
$Z \ge$ 8 has been reduced to 0.632 MeV. Very few other mean-field mass tables 
have been published, the most successful of which is the RMF table of Ref.
\cite{gtm05}, the rms error for essentially all measured nuclei being 2.1 MeV.
Also to be noted is the mass table based on the SLy4 Skyrme force; this is
limited to even-even nuclei and gives $\sigma$ = 5.1 MeV~\cite{sto03,doba04}. 
Turning to models lacking the microscopic basis of self-consistent mean-field 
models, we note that for the FRDM (``finite-range droplet 
model'')~\cite{frdm} $\sigma$ = 0.656 MeV, while for the Duflo-Zuker model~\cite{dz95} 
we have $\sigma$ = 0.360 MeV (these last two values of $\sigma$ are
both for 2149 nuclei). Only this last model does better than we do with the
model presented here, but its applicability is limited to masses: it cannot
be extended to any of the other quantities of astrophysical interest, such as 
fission barriers or the properties of the inner crust of neutron stars.

Not only is our model well adapted to the determination of these quantities,
it turns out that even with no
flexibility at all remaining for the pairing force, the spectral pairing gaps
that we find suggest that level densities in good agreement with experiment
should be obtained. Finally, we find the correct sign for the isovector 
effective mass 
splitting without introducing additional terms in the effective force. 

Much of the success of the present model must lie with the new constraint that
we have imposed, i.e., our scheme
for {\it exactly} matching the effective density-dependent contact pairing
force to realistic microscopic calculations on neutron matter. 
We suspect also that a crucial role was played by the decision to make the 
pairing force for nucleons of charge type $q$ depend only on $\rho_q$.
The modification of the collective correction (\ref{4.2}) has also contributed 
significantly to
the improvement. (Dropping Coulomb exchange certainly accounts in part for the
superiority of HFB-16 over HFB-14 and earlier models.)

It is noteworthy that widely differing neutron-matter gaps can still
correspond to the same finite-nucleus gaps, as well illustrated
by the gaps for forces BSk7, BSk8 and BSk16 shown in Fig.1, all of which
lead to comparably good mass fits. This flexibility originates in the crucial 
role that
the Skyrme force plays in passing from neutron-matter gaps to finite-nucleus
gaps. It is this flexibility which allowed us to sucessfully impose the 
additional constraint of fitting to the neutron-matter gaps. On the other hand,
it does not follow that one can start with any neutron-matter gap: for example,
when we took screened neutron-matter
gaps as our starting point we were unable to get good mass fits, no matter
how the Skyrme force was adjusted. The fact that we do get better mass fits 
while neglecting medium effects suggests that these effects might be 
compensated in finite nuclei by, for example, coupling of particles via 
surface vibrations. Alternatively, the nuclear-matter medium effects themselves
might be rather small, as indicated by recent quantum Monte Carlo 
calculations~\cite{fab05,abe07,gez07, gan08}. 

To summarize, the HFB16 mass model not only gives a   
better fit to the mass data than any other mean-field model, but is also by far
the most microscopically founded. It can thus be expected to make more reliable 
predictions of the highly neutron-rich nuclei of astrophysical interest. In 
particular, this is the first of our models well adapted to the investigation 
of a possible superfluid phase in the inner crust of neutron stars; in this 
respect it should be
realized that since the model reproduces with precision all the mass data
it must be giving a good account of surface properties, an important point
in dealing with the inhomogeneities of the crust.

\begin{acknowledgments}

We are grateful to M. Bender and T. Duguet for valuable communications. 
M. Farine is thanked for allowing us to use his HF code for semi-infinite 
nuclear matter. The financial support of the FNRS (Belgium) and the NSERC 
(Canada) is acknowledged. N.C. is grateful for the award of a Marie Curie 
Intra-European fellowship (contract number MEIF-CT-2005-024660).
\end{acknowledgments}

\appendix

\section{Relation between different formulations of the HFB method}
\label{appA}
\renewcommand{\theequation}{A.\arabic{equation}}
\setcounter{equation}{0}

{\it Computation of the HFB matrix elements.}
Our main concern in this appendix is to show how the computation of the 
oscillator matrix elements (\ref{13a}) and (\ref{13b}) is greatly facilitated 
by starting with a density functional of the general form given by 
Eq.~(\ref{17}), and working in coordinate space. First, however, we recall
the definitions and various properties of the quantities appearing in this
last equation. 

We have introduced

\noindent (i) the nucleon density,
\beqy\label{18}
\rho_q(\pmb{r}) = \sum_{\sigma=\pm 1}\rho_q(\pmb{r}, \sigma; \pmb{r}, \sigma)
\, ,
\eeqy
(ii) the kinetic-energy density (in units of $\hbar^2/2M_q$),
\beqy\label{19}
\tau_q(\pmb{r}) = \sum_{\sigma=\pm 1}\int\,{\rm d}^3\pmb{r^\prime}\,\delta(\pmb{r}-\pmb{r^\prime}) \bfdel\cdot\bfdel^\prime
\rho_q(\pmb{r}, \sigma; \pmb{r^\prime}, \sigma)
\eeqy
(iii) the spin-current density,
\beqy\label{20}
\pmb{J}_q(\pmb{r}) = -{\rm i}\sum_{\sigma,\sigma^\prime=\pm1}\int\,{\rm d}^3\pmb{r^\prime}\,\delta(\pmb{r}-\pmb{r^\prime})
\bfdel\rho_q(\pmb{r}, \sigma; \pmb{r^\prime},
\sigma^\prime) \times \mbox{\boldmath$\sigma$}_{\sigma^\prime \sigma}   \nonumber \\
={\rm i}\sum_{\sigma,\sigma^\prime=\pm1}\int\,{\rm d}^3\pmb{r^\prime}\,\delta(\pmb{r}-\pmb{r^\prime})
\bfdel^\prime\rho_q(\pmb{r}, \sigma; \pmb{r^\prime},
\sigma^\prime) \times \mbox{\boldmath$\sigma$}_{\sigma^\prime \sigma}
\eeqy
and (iv) the so-called local pairing density~\cite{doba84,doba96}
\beqy\label{27}
\tilde{\rho}_q(\pmb{r}) = \sum_{\sigma=\pm 1}
\tilde{\rho}_q(\pmb{r}, \sigma ; \pmb{r}, \sigma)   \, ,
\eeqy
where $\mbox{\boldmath$\sigma$}_{\sigma\sigma^\prime}$ denotes the Pauli spin
matrices, $\rho(\pmb{r}, \sigma; \pmb{r^\prime}, \sigma^\prime)$ and 
$\tilde{\rho}_q(\pmb{r}, \sigma ; \pmb{r}, \sigma)$ are the normal and abnormal 
density matrices respectively, expressed in so-called coordinate space 
(position $\pmb{r}$ and spin state $\sigma = \pm 1$). 
From the general definitions
of these two matrices~\cite{doba84,doba96}, it follows that the relationship
between the coordinate space representation and the discrete-basis representation 
given in Eqs.~(\ref{9a}) and (\ref{9b}), respectively, is
\bmlet
\beqy\label{22}
\rho_q(\pmb{r}, \sigma; \pmb{r^\prime}, \sigma^\prime) =
\sum_{ij(q)}\rho_{ij}\, \phi_i(\pmb{r}, \sigma)\phi^*_j(\pmb{r^\prime}, \sigma^\prime)
\eeqy
and
\beqy\label{26}
\tilde{\rho}_q(\pmb{r}, \sigma ; \pmb{r^\prime},\sigma^\prime) = 
-\sigma^\prime \sum_{ij(q)}\kappa_{ij}\, \phi_i(\pmb{r},\sigma)
\phi_j( \pmb{r^\prime},-\sigma^\prime) \, ,
\eeqy
\emlet
where $\phi_i(\pmb{r}, \sigma)$ denote the s.p. basis wave functions. 

We return now to the question of the oscillator matrix elements (\ref{13a}) and
(\ref{13b}). Remarking that $E_{\rm HFB}$ is a functional of 
$\rho_q(\pmb{r})$ (and its gradient), $\tau_q(\pmb{r})$, $\pmb{J_q}(\pmb{r})$ 
and $\tilde{\rho}_q(\pmb{r})$, which in turn depend on the matrices $\rho_{ij}$
and $\kappa_{ij}$, we have
\bmlet
\beqy\label{33a}
h_{ij}^\prime &\equiv& \frac{\partial\,E_{\rm HFB}}{\partial\rho_{ji}} \nonumber \\
&=& \int{\rm d}^3\pmb{r}\Biggl[\,\frac{\delta\,E_{\rm HFB}}{\delta\,\rho_q( \pmb{r})}
\frac{\partial\,\rho_q(\pmb{r})}{\partial\rho_{ji}} + \frac{\partial\,\mathcal{E}_{\rm HFB}(\pmb{r})}{\partial\,\tau_q( \pmb{r})}\frac{\partial\,\tau_q(\pmb{r})}{\partial\rho_{ji}} + \frac{\partial\,\mathcal{E}_{\rm HFB}(\pmb{r})}{\partial\,\pmb{J_q}( \pmb{r})}\cdot\frac{\partial\,\pmb{J_q}(\pmb{r})}{\partial\rho_{ji}} \Biggr]
\eeqy
and 
\beqy\label{33b}
\Delta_{ij} \equiv \frac{\partial\,E_{\rm HFB}}{\partial\kappa^*_{ij}}=\int{\rm d}^3\pmb{r}\,\frac{\partial\,\mathcal{E}_{\rm HFB}(\pmb{r})}
{\partial\,\tilde{\rho}_q(\pmb{r})}\,\frac{\partial\,\tilde{\rho}_q(\pmb{r})}
{\partial\kappa^*_{ij}}\, ,
\eeqy
\emlet
where we have introduced the functional derivative
\beqy\label{34a}
\frac{\delta\,E_{\rm HFB}}{\delta\,\rho_q( \pmb{r})}\equiv\frac{\partial\,\mathcal{E}_{\rm HFB}(\pmb{r})}{\partial\,\rho_q( \pmb{r})}-\pmb{\nabla}\cdot\frac{\partial\,\mathcal{E}_{\rm HFB}(\pmb{r})}{\partial\,\pmb{\nabla}\rho_q(\pmb{r})}\, .
\eeqy

Substituting Eqs.~(\ref{18}) -(\ref{20}) into Eq.~(\ref{33a}), together with
Eqs.~(\ref{22}), yields
\beqy\label{35}
h^\prime_{ij} = \sum_{\sigma,\sigma^\prime=\pm1}
\int{\rm d}^3\pmb{r}\, \phi^*_i(\pmb{r}\sigma^\prime)
h^\prime_q(\pmb{r})_{\sigma^\prime\sigma}\,\phi_j(\pmb{r}\sigma) \, ,
\eeqy
in which $h^\prime_q(\pmb{r})_{\sigma\sigma^\prime}=\sigma\sigma^{\prime} h_q'({\bf
r})_{-\sigma^{\prime}-\sigma}^*$ is the self-consistent s.p Hamiltonian,
appearing in the coordinate-space form of the HFB equations~(\ref{12b})
\beqy\label{36}
h^\prime_q(\pmb{r})_{\sigma^\prime\sigma} \equiv -\bfdel\cdot
\frac{\hbar^2}{2 M_q^*(\pmb{r})}\bfdel\, \delta_{\sigma\sigma^\prime}
+ U_q(\pmb{r}) \delta_{\sigma\sigma^\prime}
-{\rm i}\pmb{W_q}(\pmb{r}) \cdot\bfdel\times\mbox{\boldmath$\sigma$}_{\sigma^\prime\sigma}
\eeqy
where
\beqy\label{37}
\frac{\hbar^2}{2 M_q^*(\pmb{r})} =
\frac{\partial \mathcal{E}_{\rm HFB}(\pmb{r})}{\partial\tau_q(\pmb{r})}\, ,
\hskip0.5cm
U_q(\pmb{r})=\frac{\delta E_{\rm HFB}}
{\delta\rho_q(\pmb{r})}\, , \hskip0.5cm
\pmb{W}_q(\pmb{r})=\frac{\partial \mathcal{E}_{\rm HFB}(\pmb{r})}
{\partial\pmb{J}_q(\pmb{r})}  \, .
\eeqy

To reduce Eq. (\ref{33b}) in the same way requires that we make use of
Eqs. (\ref{27}) and (\ref{26}), and also the expression
\beqy\label{38}
 \phi_{_{\bar i}}(\pmb{r},\sigma) \equiv -i\sigma_y \phi_i^*(\pmb{r},\sigma)
= -\sigma \phi_i^*(\pmb{r},-\sigma)
\eeqy
for the time-reversed conjugate $\phi_{_{\bar i}}(\pmb{r},\sigma)$ of the s.p.
state $\phi_i(\pmb{r},\sigma)$. In
this way we find
\beqy\label{39}
\Delta_{ij} = \sum_{\sigma=\pm1} \int{\rm d}^3\pmb{r}\,
\phi^*_i(\pmb{r}\sigma)\Delta_q(\pmb{r})\phi_{_{\bar j}}(\pmb{r}\sigma) \, ,
\eeqy
in which $\Delta_q(\pmb{r})$ is the self-consistent local pairing potential,
appearing in the coordinate-space form of the HFB equations~(\ref{12b})
\beqy\label{40}
\Delta_q(\pmb{r}) \equiv \frac{\partial \mathcal{E}_{\rm HFB}(\pmb{r})}
{\partial \tilde{\rho}_q(\pmb{r})} \, ;
\eeqy
this must be real if time-reversibility holds, since $\tilde{\rho}_q(\pmb{r})$
will then be real~\cite{doba84,doba96}. 

Explicit expressions for the fields appearing in Eqs.~ (\ref{37}) and 
(\ref{40}) are as follows. 
\beqy
\label{A7}
\frac{\hbar^2}{2 M_q^*} = \frac{\partial \mathcal{E}_{\rm HFB}}{\partial\tau_q} = \frac{\hbar^2}{2 M_q} &+& \frac{1}{4} t_1 \Biggl[ \left(1+ \frac{1}{2} x_1\right) \rho 
- \left( \frac{1}{2} + x_1\right)\rho_q\Biggr]  \nonumber \\
 &+& \frac{1}{4} t_2 \Biggl[ \left(1+ \frac{1}{2} x_2\right) \rho + \left( \frac{1}{2} + x_2\right)\rho_q\Biggr] \, ,
\eeqy
\beqy\label{A8}
U_q&=&\frac{\delta E_{\rm HFB}}{\delta\rho_q} = \frac{\partial \mathcal{E}_{\rm HFB}}{\partial\rho_q} - \pmb{\nabla}\cdot\frac{\partial \mathcal{E}_{\rm HFB}}{\partial (\pmb{\nabla}\rho_q)} \nonumber \\
&=& t_0\Biggl[\left(1+ \frac{1}{2} x_0\right)\rho - \left(\frac{1}{2} +x_0\right)\rho_q\Biggr] \nonumber \\
&+&\frac{1}{4} t_1 \Biggl[\left(1+ \frac{1}{2} x_1\right) \left(\tau - \frac{3}{2}\nabla^2\rho\right) - \left(\frac{1}{2} +x_1\right)\left(\tau_q - \frac{3}{2}\nabla^2\rho_q\right) \Biggr]  \nonumber \\
&+&\frac{1}{4} t_2 \Biggl[\left(1+ \frac{1}{2} x_2\right) \left(\tau + \frac{1}{2}\nabla^2\rho\right) + \left(\frac{1}{2} +x_2\right)\left(\tau_q + \frac{1}{2}\nabla^2\rho_q\right) \Biggr]  \nonumber \\
&+&\frac{1}{12} t_3 \Biggl[\left(1+ \frac{1}{2} x_3\right) (2+\gamma)\rho^{\gamma+1} - \left(\frac{1}{2} +x_3\right)\left(2\rho^\gamma \rho_q+\gamma \rho^{\gamma-1} \sum_{q^\prime=n,p} \rho_{q^\prime}^2 \right) \Biggr]  \nonumber \\
&-&\frac{1}{2} W_0 \left(\pmb{\nabla}\cdot \pmb{J} + \pmb{\nabla}\cdot \pmb{J_q}\right)+ \delta_{q,p} V^{\rm Coul}+\frac{1}{4} \sum_{q^\prime=n,p}\frac{\partial v^{\pi q^\prime}}{\partial \rho_q}\, \tilde{\rho}_{q^\prime}^2 
\eeqy
and
\beqy\label{A9}
\pmb{W_q}=\frac{\partial \mathcal{E}_{\rm HFB}}{\partial \pmb{J_q}}=\frac{1}{2} W_0 \pmb{\nabla} (\rho+\rho_q) -\frac{1}{8} (t_1 x_1 + t_2 x_2) \pmb{J} + \frac{1}{8} (t_1-t_2) \pmb{J_q} \, .
\eeqy
Note that for the Coulomb field term $\delta_{q,p} V^{\rm Coul}$ appearing in
Eq.~(\ref{A8}) we have taken the electrostatic potential given by 
Eq.~(\ref{A4}), which, in view of the finite proton size, is not given exactly 
by the functional derivative
of the Coulomb energy $(\delta E_{\rm Coul})/(\delta \rho_p)$.  

The local pairing field, defined in Eq.~(\ref{40}), is given by 
\beqy\label{A10}
\Delta_q=\frac{\partial \mathcal{E}_{\rm HFB}}{\partial \tilde{\rho}_q}=\frac{1}{2}v^{\pi q} [\rho_n,\rho_p] \tilde{\rho}_q \, .
\eeqy
Let us point out that the pairing functional $\mathcal{E}_{\rm pair}$ 
contributes not only to this pairing field but also, through its
dependence on the nucleon density, to the s.p. Hamiltonian
$h^\prime_q(\pmb{r})_{\sigma\sigma^\prime}$ as the last term of Eq. (\ref{A8}),
which is essentially a rearrangement term, corresponding to the second term of
Eq.~(\ref{15b}). (All the other rearrangement field terms, corresponding to
the first term of Eq.~(\ref{15b}), appear in Eq. (\ref{A8}), and can be 
identified by their coefficient $t_3\gamma$.)   

{\it Transformation from discrete-basis to coordinate-space formulations of HFB
equations.} It is convenient now to
point out that the HFB equations in coordinate space can be obtained 
directly from the standard formulation of Refs.~\cite{mang75,rs80}, i.e.,
from the corresponding equations in discrete-basis form~(\ref{12}), by 
substituting in the latter Eqs.~(\ref{35}) and (\ref{39}), and invoking
time-reversibility.

This leads for each nucleon
species $q$ to the coordinate-space form of these equations, as given by 
Refs.~\cite{doba84,doba96}, thus 
\beqy\label{12b}
\sum_{\sigma^\prime=\pm1}
\begin{pmatrix} h^\prime_q(\pmb{r} )_{\sigma \sigma^\prime} -\lambda_q\, \delta_{\sigma \sigma^\prime} & \Delta_q(\pmb{r}) \delta_{\sigma \sigma^\prime} \\ \Delta_q(\pmb{r}) \delta_{\sigma \sigma^\prime} & -h^\prime_q(\pmb{r})_{\sigma \sigma^\prime} 
+ \lambda_q\, \delta_{\sigma \sigma^\prime} \end{pmatrix}\begin{pmatrix} 
\psi^{(q)}_{1i}(\pmb{r},\sigma^\prime) \\ \psi^{(q)}_{2i}(\pmb{r},\sigma^\prime) \end{pmatrix} =
E_i \begin{pmatrix} \psi^{(q)}_{1i}(\pmb{r},\sigma) \\ \psi^{(q)}_{2i}(\pmb{r},\sigma) \end{pmatrix} \, ,
\eeqy
where the s.p. Hamiltonian $h^\prime_q(\pmb{r} )_{\sigma \sigma^\prime}$ and 
pairing field $\Delta_q(\pmb{r})$ for the forces~(\ref{1}) and (\ref{2}) are as 
given in Appendix~\ref{appA} (note that the rearrangement terms are included
here). The upper and lower components of the Bogoliubov quasiparticle wave 
function, 
denoted respectively by $\psi^{(q)}_{1i}(\pmb{r},\sigma)$ and 
$\psi^{(q)}_{2i}(\pmb{r},\sigma)$, are related to the $U$ and $V$ matrices 
introduced in Eq.~(\ref{7}) by 
\beqy
\label{13}
\psi^{(q)}_{1i}(\pmb{r},\sigma) = \sum_{j(q)} U_{ji}\, \phi_j(\pmb{r},\sigma)\, , \hskip 0.5cm \psi^{(q)}_{2i}(\pmb{r},\sigma) = \sum_{j(q)} V_{ji}\, \phi_{_{\bar j}}(\pmb{r},\sigma) \, ;
\eeqy
note the appearance of the time-reversed conjugate state in the latter 
expression (Eq.~(6) of Ref~\cite{bhr03} is wrong).

Using Eqs.~(\ref{13}), (\ref{22}), (\ref{26}) and  
the assumed time-reversal invariance, we have the identities
\beqy
\rho_q(\pmb{r}, \sigma; \pmb{r^\prime}, \sigma^\prime) =
\sum_{i(q)}\psi^{(q)}_{2i}(\pmb{r}, \sigma)\psi^{(q)}_{2i}(\pmb{r^\prime}, \sigma^\prime)^* 
\eeqy
and
\beqy
\tilde{\rho}_q(\pmb{r}, \sigma; \pmb{r^\prime}, \sigma^\prime) =
-\sum_{i(q)}\psi^{(q)}_{2i}(\pmb{r}, \sigma)
\psi^{(q)}_{1i}(\pmb{r^\prime}, \sigma^\prime)^*=-
\sum_{i}\psi^{(q)}_{1i}(\pmb{r}, \sigma)\psi^{(q)}_{2i}(\pmb{r^\prime},
\sigma^\prime)^* \, .
\eeqy 
Applying the properties of the $U$ and $V$ matrices, as given for
example by Eqs.~(7.5) of Ref.~\cite{rs80}, to the definition~(\ref{13}) it can 
be easily checked that the quasiparticle wavefunction satisfies the completeness 
relations given by Eqs.~(2.20a) and (2.20b) of Ref.~\cite{doba84}.

\section{HFB equations in uniform matter}
\label{appC}
\renewcommand{\theequation}{B.\arabic{equation}}
\setcounter{equation}{0}

In a uniform system, it is natural
to replace the oscillator basis by a plane-wave basis,
\beqy\label{C1}
\phi_k(\pmb{r},\sigma) \equiv \frac{1}{\sqrt{\cal V}}\, \exp\left({\rm i} \pmb{k}\cdot\pmb{r}\right) \chi(\sigma) \, ,
\eeqy
where $\chi(\sigma)$ is the Pauli spinor and $\cal V$ is the normalization volume. It is easily seen from Eq.~(\ref{35}) 
that the s.p. Hamiltonian $h^\prime_q(\pmb{r} )_{\sigma \sigma^\prime}$ is 
diagonal in this basis,
\beqy\label{C2}
h^\prime_{kl}=\varepsilon^{(q)}_k\, \delta_{kl} \, ,
\eeqy
 with 
\beqy\label{C3}
\varepsilon^{(q)}_k =\frac{\hbar^2 k^2}{2 M_q^*} +U_q \, .
\eeqy
Also, since the pairing field $\Delta_q(\pmb{r})$ must be independent of 
${\pmb{r}}$ in a uniform system it follows from Eq.~(\ref{39}) that all matrix 
elements $\Delta_{lk}$ vanish unless $l = {\bar k}$, i.e.,  
\beqy\label{C4}
\Delta_{lk} =\delta_{l{\bar k}}\, \Delta_q  = -\Delta_{kl}\,
\eeqy
where $\Delta_q$ is a constant for a given uniform system at a given density,
given by Eq. (\ref{A10}).
The assumption that the pairing force acts only between states of the s.p. 
Hamiltonian that are the time-reversal of each other constitutes the essence of
the BCS method (note that this approximation differs from that discussed in Ref.~\cite{doba96}), an
approximation to the HFB method that is often convenient for finite nuclei,
but we have shown here that for uniform systems the HFB method reduces exactly
to the BCS method. Furthermore, for infinite but inhomogeneous nuclear matter,
the weaker the departure from homogeneity the better the BCS method will
approximate the HFB method; this means that in the inner crust of neutron stars
the BCS approximation will be better the deeper the layer in question.

More precisely, by inspecting Eq.~(\ref{39}) and by remembering that only s.p. 
states close to the Fermi level contribute to pairing correlations, it can be 
seen that in general the BCS approach is justified whenever the pairing field 
is slowly varying in the spatial domain for which the s.p. wavefunctions for 
states around the Fermi level, are non-vanishing. This condition is usually 
fulfilled for strongly bound nuclei (for which the chemical potential 
$\lambda_q$ lies deep inside the s.p. potential well) since the pairing field 
$\Delta_q(\pmb{r})$ is typically more or less constant in the nuclear interior,
where the bound s.p. wavefunctions take their largest values. In contrast, for 
weakly bound nuclei, pairing correlations involve not only bound states but 
also states from the continuum. Since the pairing field vanishes outside the 
nucleus, it follows that it will be varying rapidly in a region where the s.p. 
states that contribute to pairing are still strong, whence the BCS 
approximation can be expected to break down for such nuclei. 

Solving now the HFB equations~(\ref{12}) by using Eqs.~(\ref{C2}) and 
(\ref{C4}) yields
\bmlet
\beqy
\label{C5a}
E^{(q)}_k = \sqrt{(\varepsilon^{(q)}_k-\lambda_q)^2+\Delta_q^2} \, ,
\eeqy
\beqy\label{C5b}
U^{(q)}_{k k}=U^{(q)}_{\bar k\bar k} = \frac{1}{\sqrt{2}}\left(1+\frac{\varepsilon^{(q)}_k-\lambda_q}{E^{(q)}_k}\right)^{1/2} 
\eeqy
and
\beqy\label{C5c}
V^{(q)}_{k \bar k}=-V^{(q)}_{\bar k k} = \frac{1}{\sqrt{2}}\left(1-\frac{\varepsilon^{(q)}_k-\lambda_q}{E^{(q)}_k}\right)^{1/2} \, ,
\eeqy
\emlet
making use of the properties of the $U$ and $V$ matrices. Using Eq.~(\ref{13}), (\ref{C5b}) and (\ref{C5c}) it can be seen that the 
quasiparticle wavefunction in uniform nuclear matter reduces to 
\beqy
\psi^{(q)}_{1k}(\pmb{r},\sigma) = U^{(q)}_{k k}\, \phi_k(\pmb{r},\sigma)  \, , \hskip0.5cm \psi^{(q)}_{2k}(\pmb{r},\sigma)=V_{\bar k k}\, \phi_k(\pmb{r},\sigma) \, ,
\eeqy
where $\phi_k(\pmb{r},\sigma)$ is given by Eq.~(\ref{C1}).

Substituting Eqs.~(\ref{C5b}) and (\ref{C5c}) in Eqs.~(\ref{9a}) and (\ref{9b}) leads to the familiar expressions 
of the normal and abnormal density matrices
\beqy\label{C6}
\rho_{kl}= (V_{k \bar k})^2\, \delta_{kl}\, , \hskip 0.5cm 
\kappa_{kl}=V_{k \bar k} U_{k k}\, \delta_{\bar k l} \, .
\eeqy
Using Eq.~(\ref{18}) and (\ref{22}) together with (\ref{C1}) and (\ref{C5c}), 
the nucleon density $\rho_q(\pmb{r})\equiv\rho_q$ is given by
\beqy
\label{C7}
\rho_q = \frac{1}{4\pi^2} \left(\frac{2 M_q^*}{\hbar^2}\right)^{3/2} \int_{U_q}^{+\infty}{\rm d}\varepsilon \sqrt{\varepsilon}\left(1-\frac{\varepsilon-\lambda_q}{\sqrt{(\varepsilon-\lambda_q)^2+\Delta_q^2}}\right) \, .
\eeqy
Likewise from Eqs.~(\ref{C6}) and (\ref{11b}) we find the ``BCS gap equation''
\beqy
\label{C8}
\Delta_q=- \frac{1}{8\pi^2} \left(\frac{2 M_q^*}{\hbar^2}\right)^{3/2} v^{\pi\, q}[\rho_n,\rho_p] \, \Delta_q \,\int_\Lambda {\rm d}\varepsilon 
\frac{\sqrt{\varepsilon}}{\sqrt{(\varepsilon-\lambda_q)^2+\Delta_q^2}} \, .
\eeqy
The integral is taken inside the subspace $\Lambda$ introduced to regularize
the ultra-violet divergences, arising from the zero range of the pairing interaction. 
These divergences are removed by imposing a cutoff, either in the q.p. energy spectrum 
or in the s.p. energy spectrum. In the latter case, different prescriptions have been 
employed in the literature: (i) $\varepsilon < \varepsilon_{\Lambda}$, (ii) $\varepsilon < U_q+\varepsilon_{\Lambda}$, 
(iii) $\varepsilon < \lambda_q+\varepsilon_{\Lambda}$, 
(iv) $\lambda_q-\varepsilon_{\Lambda}<\varepsilon < \lambda_q+\varepsilon_{\Lambda}$ where 
$\varepsilon_\Lambda$ is a constant. The fixed cutoff (i) is the only choice which implies that 
the gap equations depend explicitly on the s.p. potential $U_q$. Since a density dependent
pairing force affects directly the s.p. energies via the rearrangement potential (see Eq.~(\ref{A8})), 
the gap equations~(\ref{C8}) will involve not only the pairing strength $v^{\pi\, q}[\rho_n,\rho_p]$ but 
also its partial derivatives. Note that since the integrand in the integral appearing in 
Eq.~(\ref{C8}) takes significant values only in the vicinity of the Fermi level, the cutoff prescriptions 
(i) and (ii) lead to essentially zero pairing gaps $\Delta_q \simeq 0$ for densities such that 
$\varepsilon_{\Lambda}<\lambda_q$ and $\varepsilon_{\Lambda}<\mu_q$ respectively. 
A peculiar consequence is that the pairing gap may be vanishingly small while the pairing force is not. 
We have encountered such a situation for the Skyrme force SLy4 with the pairing force of Ref.~\cite{doba04} 
using the cutoff prescription (i), as can be seen by comparing Figs.~\ref{fig_deltan_bcs} and \ref{fig_vpiq}.

\newpage

\begin{table}
\centering
\caption{Parameters of the analytical fit~(\ref{eq.fit.gap}) for the BCS 
pairing gap in neutron matter extracted from Fig.~7 of
Ref.~\cite{lom01} (the unit of length is fermi and the unit of energy is MeV).}
\label{tab1} 
\vspace{.5cm}
\begin{tabular}{|c|c|c|c|c|}
\hline $\Delta_0$ & $k_1$ & $k_2$ & $k_3$ & $k_{\rm max}$ \\
\hline 910.603 & 1.38297 & 1.57068 & 0.905237 & 1.57 \\ \hline \end{tabular}
\end{table}

\vspace{1.0cm}

\begin{table}
\centering
\caption{Force BSk16: lines 1-10 show the Skyrme parameters, lines 11-14 
the pairing parameters and the last four lines the Wigner parameters
(see Section \ref{sect.fits} for further details).} 
\label{tab2}
\vspace{.5cm}
\begin{tabular}{|c|c|}
\hline
  $t_0$ {\scriptsize [MeV fm$^3$]}   & -1837.23  \\
  $t_1$ {\scriptsize [MeV fm$^5$]}   & 383.521   \\
  $t_2$ {\scriptsize [MeV fm$^5$]}   & -3.41736 \\
  $t_3$ {\scriptsize [MeV fm$^{3+3\gamma}$]}  & 11523.0  \\
  $x_0$                              &  0.432600  \\
  $x_1$                              & -0.824106 \\
  $x_2$                              & 44.6520  \\
  $x_3$                              &  0.689797  \\
  $W_0$ {\scriptsize [MeV fm$^5$]}   &  141.100    \\
  $\gamma$                           &  0.3   \\
  $f_{n}^-$  &  1.06  \\
  $f_{p}^+$  &  0.99  \\
  $f_{p}^-$  & 1.05   \\
  $\varepsilon_{\Lambda}$ {\scriptsize [MeV]}  &  16.0    \\
  $V_W$ {\scriptsize [MeV]}           & -2.60   \\
  $\lambda$                           & 240    \\
  $V_W^{\prime}$ {\scriptsize [MeV]}  & 0.70   \\
  $A_0$                               &32    \\
 \hline
\end{tabular}
\end{table}

\begin{table}
\centering
\caption{Parameters of collective correction for model HFB-16 (see Section 
\ref{sect.fits} for further details).}
\label{tab3}
\vspace{.5cm}
\begin{tabular}{|c|c|}
\hline
$b$  (MeV)&  0.8\\
$c$ & 10 \\
$d$ (MeV) & 2.6 \\
$l$ & 10 \\
$\beta_2^0$ & 0.1\\
 \hline
\end{tabular}
\end{table}

\begin{table}
\centering
\caption{Rms ($\sigma$) and mean ($\bar{\epsilon}$) deviations 
between data and predictions for model HFB-16; for convenience we also show 
models HFB-15 \cite{gp08}, HFB-14 \cite{sg07} and HFB-8 \cite{ms04}. The 
first pair of lines 
refers to all the 2149 measured masses $M$, the second pair to the masses 
$M_{nr}$ of the subset of 185 neutron-rich nuclei with $S_n \le $ 5.0 MeV, the 
third pair to the neutron separation energies $S_n$ (1988 measured values), the
fourth pair to beta-decay energies $Q_\beta$  (1868 measured values) and
the fifth pair to charge radii (782 measured values). The last line shows
the calculated neutron-skin thickness of $^{208}$Pb for these models.}
\label{tab4}
\vspace{.5cm}
\begin{tabular}{|c|cccc|}
\hline
&HFB-16&HFB-15& HFB-14 & HFB-8  \\
\hline
$\sigma(M)$ {\scriptsize [MeV]} &0.632&0.678&0.729  &0.635   \\
$\bar{\epsilon}(M)$ {\scriptsize [MeV]}&-0.001&0.026 &-0.057 &0.009  \\
$\sigma(M_{nr})$ {\scriptsize [MeV]}&0.748 &0.809&0.833  &0.838      \\
$\bar{\epsilon}(M_{nr})$ {\scriptsize [MeV]}&0.161&0.173&0.261&-0.025\\
$\sigma(S_n)$ {\scriptsize [MeV]}&0.500 &0.588 &0.640 &0.564\\
$\bar{\epsilon}(S_n)$ {\scriptsize [MeV]}&-0.012   &-0.004 & -0.002&0.013\\
$\sigma(Q_\beta)$ {\scriptsize [MeV]}&0.559  &0.693&0.754&0.704\\
$\bar{\epsilon}(Q_\beta)$ {\scriptsize [MeV]}&0.031 &0.024&0.008&-0.027 \\
$\sigma(R_c)$ {\scriptsize [fm]}&0.0313 &0.0302&0.0309 &0.0275\\
$\bar{\epsilon}(R_c)$ {\scriptsize [fm]}&-0.0149&-0.0108&-0.0117&0.0025\\
$\theta$($^{208}$Pb) {\scriptsize [fm]}&0.15  &0.15 &0.16&0.12\\
\hline
\end{tabular}
\end{table}

\begin{table}
\centering
\caption{Macroscopic parameters for force BSk16 (for convenience we also show
forces BSk 15 \cite{gp08} and BSk14 \cite{sg07}). The first twelve lines refer 
to infinite nuclear matter, the last two to semi-infinite nuclear matter.
See Section \ref{sect.fits} for further details.}
\label{tab5}
\vspace{.5cm}
\begin{tabular}{|c|ccc|}
\hline
&BSk16&BSk15&BSk14\\
\hline
$a_v$ {\scriptsize [MeV]}&-16.053&-16.037&-15.853 \\
$\rho_0$ {\scriptsize [fm$^{-3}$]}&0.1586&0.1589    &0.1586 \\
$J$ {\scriptsize [MeV]}&30.0&30.0 &30.0    \\
$M^*_s/M$ &0.80& 0.80&0.80   \\
$M^*_v/M$&0.78& 0.77   & 0.78       \\
$K_v$ \scriptsize [MeV]&241.6& 241.5  & 239.3 \\
$L$ \scriptsize [MeV]&34.87 &33.60& 43.91 \\
$G_0$ &-0.65&-0.67&-0.63 \\
$G_0^{'}$&0.51&  0.54  &0.51  \\
$G_1$&1.52   &  1.47 &1.49 \\
$G_1^{'}$ &0.44 &  0.41  &0.44 \\
$\rho_{frmg}/\rho_0$&1.24&1.24 &1.24 \\
$a_{sf}$ \scriptsize [MeV]&17.8 &17.7&17.6 \\
$Q$ \scriptsize [MeV]&39.0&39.7&35.0 \\
\hline
\end{tabular}
\end{table}

\begin{table}
\centering
\caption{Rms and mean differences between predictions for highly
neutron-rich nuclei (4.0 MeV $\ge S_n \ge 0$ and $26 \le Z \le 110$)
given by different pairs of mass models.
Mean differences are shown in parentheses. }
\label{tab6}
 \vspace{.5cm}
 \tabcolsep=.5cm
 \begin{tabular}{|c|ccc|}
 \hline
&$M$ &$S_n$& $Q_{\beta}$\\
\hline
HFB-15 - HFB-16&1.356 (0.977)&0.378 (-0.033) & 0.520 (0.092)\\
HFB-14 - HFB-16&3.230(-2.465)&0.392 (0.136)&0.595 (-0.343)\\
HFB-8 - HFB-16&1.581 (0.319)&0.581 (-0.099)&0.892 (0.198)\\
 \hline
 \end{tabular}
 \end{table}
\end{document}